\documentclass[twocolumn,trackchanges]{aastex631}

\usepackage{amsmath}
\usepackage{float}
\usepackage{scalerel}
\usepackage{lipsum}
\usepackage {hyperref}

\shorttitle{Distances, Radial Distribution and total number of Galactic SNRs}
\shortauthors{Ranasinghe, S. \& Leahy, D.}

\begin{document}

\title{Distances, Radial Distribution and Total Number of Galactic Supernova Remnants}

\author[0000-0002-9559-3827]{S. Ranasinghe}
\affiliation{Department of Physics and Astronomy, University of Calgary, 2500 University Dr NW, Calgary, AB, T2N 1N4, Canada}

\author[0000-0002-4814-958X]{D. Leahy}
\affiliation{Department of Physics and Astronomy, University of Calgary, 2500 University Dr NW, Calgary, AB, T2N 1N4, Canada}


\begin{abstract}
 We present a table of 215 SNRs with distances. 
New distances are found to SNR G$51.26+0.11$ of $6.6 \pm 1.7$ kpc  using HI absorption spectra   
and to 5 other SNRs using maser/molecular cloud associations.
We recalculate the distances and errors to all SNRs using a consistent rotation curve and provide errors where they were not previously estimated. 
This results in a significant distance revisions for 20 SNRs. 
Because of observational constraints and selection effects, there 
is an apparent deficit of observed number of Galactic supernova remnants (SNRs). 
To investigate this, we employ two methods. The first method applies correction factors for the selection effects to derive the radial density distribution. 
The second method compares functional forms for the SNR surface density and selection function against the data 
to find which functions are consistent with the data. 
The total number of SNRs in the Galaxy is $\sim3500$ (Method 1) or in the range $\sim2400$ to $\sim5600$ (Method 2).   
We conclude that the current observed number of SNRs is not yet complete enough to give a well-determined total SNR number or 
radial density function. 
\end{abstract}



\section{Introduction} \label{sec:intro}

Supernova remnants (SNRs) are an important area of study, as they give insight in to the evolution of the interstellar medium (ISM) and galaxies. 
To better understand our galaxy, not only the study of individual SNRs is important but also an accurate count of Galactic SNRs is essential. 
Currently there are 294 known Galactic SNRs \citep{2019GreenCat} and there is an apparent deficit between the number of observed and expected Galactic SNRs based on supernovae rates.\\
\indent  If the Supernova (SN) rate of the Milky way galaxy is one per $40 \pm 10$ yr \citep{1994Tammann} and we assume the mean lifetime of radio SNRs to be $\sim60,000$ yr \citep{1994Frail}, the expected number of Galactic SNRs is of order 1500. 
This leaves the majority of SNRs yet to be discovered. 
The probable reason for the deficit is that the surveys searching for radio SNRs have selection effects \citep{1991Green, 1998CaseBhat}.  
Mainly they overlook faint SNRs as well as young compact SNRs due to the limitations on resolution and sensitivity. 
Identifying large faint remnants remains a difficult task because of their low surface brightness and complex structure. \\
\indent Identifying young compact SNRs is challenging as well, since they lack a shell-like morphology and are difficult to distinguish from extragalactic sources. 
Based on both statistics of known Galactic SNRs and SN rate, \cite{2021Ranasinghe} stated the expected number of young compact SNRs in our galaxy is $\sim15 -20$. 
They found 2 compact SNRs in the region covered by the THOR survey \citep{2016Beuther, 2020Wang}, which was half the expected value. 
However, this number was found to be consistent with the THOR sensitivity limit. 
Even with a liberal estimate for these compact SNRs, it falls short to compensate for the deficit. \\
\indent The main goal of this work is to examine the radial distribution of Galactic SNRs and the affect of observational selection effects. 
To this end we employ two methods. 
First, we use a simple empirical method for the correction of observational selection effects as previously investigated by \cite{1974Kodaira} \&  \cite{1989LeahyXinji} with a smaller sample. 
The second method uses fitting of functional forms to the observed data to obtain functions that best describe the radial distribution.\\
\indent  In order to investigate the radial distribution of Galactic SNRs, accurate distance estimates for the SNRs are crucial. 
There are a considerable number of SNRs that have distances estimated using the $\Sigma$-D relation (surface brightness ($\Sigma$)- physical diameter (D) relation). 
However, there is a large scatter in the $\Sigma$-D distribution for Galactic SNRs and the physical diameter varies by about an order of magnitude \citep{1991Green, 2005Greenmem}. 
Hence, for this work we utilize distances obtained by more reliable methods (e.g. HI absorption spectra and molecular cloud (MC) interactions).\\
\indent  In Section \ref{Method}, the compilation of the SNR distances are presented. 
Description of the two methods used to estimate the SNR surface density distribution and their results are given in Section \ref{Results}. The discussion and conclusion are presented in Sections \ref{Discussion} \& \ref{conclusion}, respectively. 

\section{Compilation of SNR Distances}  \label{Method}

\indent To obtain the source list, we use the catalogue presented by \cite{2019GreenCat}\footnote{\url{http://www.mrao.cam.ac.uk/surveys/snrs/}}. 
Apart from this catalogue, we use the catalogue presented by \cite{2012Ferrand}\footnote{\url{http://www.physics.umanitoba.ca/snr/SNRcat}} to search for remnants with reliable distances. 
Furthermore, we performed a literature search for newly identified SNRs. \\
\indent For SNRs with no published distances, if there are molecular cloud or maser associations, we infer the distances from those.  
The literature  distances have been estimated using different rotation curves, where the distance to the Galactic center (GC), $R_{0}$ ranging from 7.6 to 8.5 kpc and the orbital velocity of the Sun, $V_0$ ranging from 200 to 245 km s$^{-1}$. 
In this work, for consistency we recalculate the distances using the values presented by \cite{2014ReidRC} of $R_{0} = 8.34 \pm 0.16$ and $V_0 = 241 \pm 8$ km s$^{-1}$ along with their ``Univ" parameters to estimate a distance where an associated velocity is given. 
The errors of the distances were recalaculated as described by \cite{2018RanasMNRAS}.

\subsection{Distances to Galactic SNRs} \label{Galdist}

\indent \cite{2019GreenCat} gives the current number of Galactic SNRs as 294 while the catalogue presented by \cite{2012Ferrand} gives the current number as 383. 
These 383 SNRs include uncertain SNRs and candidate SNRs in \cite{2019GreenCat}. 
The uncertain SNRs need further observations to confirm their nature and we include them in this study, while the SNR candidates are not. Many studies have produced SNR candidates (e.g. \cite{2017Anderson}, \cite{2019Hurley}). However, some of these candidates may not be SNRs and add contamination to the sample of SNRs. Therefore, as a caution we exclude them from this study. \\
\indent Out of the 294 SNRs in Green's catalogue, 192 SNRs have distances estimated using reliable methods. 
The list of the sources is given in Table \ref{tab:1} with revised distances where relevant. 
There are 17 uncertain SNRs with distances that are found in the catalogue presented by \cite{2012Ferrand} (Table \ref{tab:1} sources with $\dagger$). \\
\indent Furthermore, a literature search produced 6 newly identified SNRs (Table \ref{tab:1} sources with $\ddagger$), bringing the total number of sources in this sample to 215. 

\subsubsection{New Distance Estimate for SNR G$51.26+0.11$}

\begin{figure}
\centering
\includegraphics[width=\columnwidth]{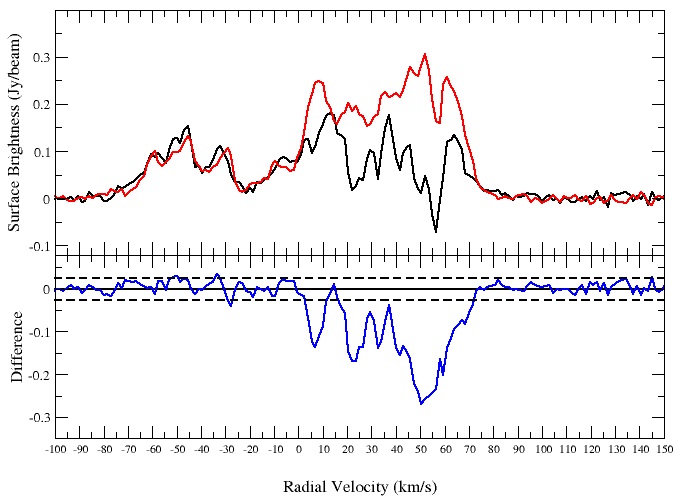}
\caption {HI spectrum of the SNR G$51.26+0.11$. Top: HI emission spectra (source: black; background: red). 
Bottom: source – background difference: (blue). The $\pm2\sigma$ noise level of the difference is shown by the dashed lines.}
\label{fig:2}
\end{figure}

\indent G$51.21+0.11$ was given as a SNR candidate by \cite{2017Anderson}. 
With a $\sim11\farcm$3 radius, it is located $30\arcmin$ from the SNR G$51.04+0.07$ and reclassified as G$51.26+0.11$ by \cite{2018Dokara}.\\
\indent The HI spectra for the SNR G$51.21+0.11$ were constructed 
with data from the THOR data release 2 \citep{2020Wang}. \\
\indent  Examination of the HI spectra (Figure \ref{fig:2} and individual HI channel maps, shows absorption up to the tangent point ($\sim70$ km s$^{-1}$ or $5.2$ kpc). 
Therefore, the SNR is located beyond the tangent point. 
HI channel maps do not show absorption at $15$ or $37$ km s$^{-1}$. 
Because we do not see absorption at $37$ km s$^{-1}$, the upper limit is the corresponding distance of $7.7$ kpc. 
We place the SNR at a distance of $6.6 \pm 1.7$kpc.

\subsection{Revised Distances to SNRs}

\indent The revised distance estimates of SNRs were done using a consistent rotation curve \citep{2014ReidRC}  throughout this work. 
We present the distance estimates for one SNR (G$5.5+0.3$) with a MC association and four SNRs (G$341.9-0.3$, G$342.0-0.2$, G$343.1-0.7$  and  G$354.8-0.8$) with maser associations. 
The analyses of the distances to 9 SNRs were re-done with a combination of either new information or evidence. 
These distance estimates are a significant improvement from previously known values. 
The positions of the Galactic SNRs shown in figure \ref{fig:1} includes the revised distances from Table \ref{fig:1}\\

G$5.5+0.3$: There were no previous distance estimated for this SNR. 
\cite{2009Liszt} suggested a possible CO association with the SNR at $10$ - $15$ km s$^{-1}$. 
We take the average 12.5 km s$^{-1}$ to estimate the near distance of  $3.2\pm0.7$ kpc, consistent with \cite{2009Liszt}.

G$11.2-0.3$: For SNR $11.2-0.3$, the HI absorption spectrum was shown by \cite{1985Becker}. 
With no HI absorption up to the tangent point, \cite{2004GreenDA} recalculated the distance to be $4.4$ kpc. 
However, \cite{2016Kilpatrick} reported velocity-broadened molecular emission toward the remnant at an average of $32.5$ km s$^{-1}$. 
This is consistent with the HI data and we place the SNR at the near distance of $3.7 \pm 0.2$ kpc.

G$53.6-2.2$: From HI observations \cite{1998Giacani} adopted a systematic velocity of $+ 27$ km s$^{-1}$ for the SNR but favoured the near kinematic distance $2.3 \pm 0.8$ kpc. 
However, \cite{2018Shan} using the optical extinction - distance relation obtained a distance lower limit of $5.3$ kpc. 
We place the SNR at the far distance of $7.8 \pm 0.6$ kpc.

G$311.5-0.3$:  \cite{1975Caswell} analysed HI absorption to place the SNR at $d > 6.6$ kpc. 
Examination of their spectrum (their Figure 6), shos that at $-10$ km s$^{-1}$ there is no absorption, rather clear absorption at $\sim 0$ km s$^{-1}$. 
Therefore, we place the SNR at the far distance of $10.3 \pm 0.5$ kpc which corresponds to the velocity of $-10$ km s$^{-1}$.

G$312.4-0.4$: \cite{2003Doherty} presented HI absorption spectra for the SNR and gave a lower limit distance of $6$ kpc. 
The tangent point velocity in this direction is $\sim-65$ km s$^{-1}$. 
Comparing the point source spectrum and the SNR HI absorption spectrum (their Figure 7 \& 8), 
the SNR absorption doesn't show up to the tangent point (the point source shows absorption up to the tangent point).
 We believe it's likely that the SNR is located at a distance of 3.5 kpc corresponding to $-50$ km s$^{-1}$. 

G$316.3+0.0$:  The HI absorption spectrum presented by \cite{1975Caswell} (their Figure 9) clearly indicates absorption up to the tangent point. 
The SNR lies beyond the corresponding distance of $6$ kpc. 
However, there is no absorption present at $\sim-40$ km s$^{-1}$ which leads to the conclusion that the SNR is located at less than the corresponding distance. Therefore, we place the SNR at a distance of $9.4 \pm 0.4$ kpc.

G$337.2-0.7$:  \cite{2006Rakowski} reported the SNR lies between a distance of $2.0 \pm 0.5$ and $9.3 \pm 0.3$ kpc. Their Figure 4 shows the absorption spectra for the HII region G$337.1-0.2$ and the SNR, with both objects showing absorption up to $-116$ km s$^{-1}$. 
This is the tangent point velocity and because the HII region is located at a distance of 11 kpc \citep{1999Corbel}, the SNR must be located beyond the tangent point. 
Examining both spectra, it is seen that the HII region shows absorption at $-100$ km s$^{-1}$ while the SNR does not. We believe the SNR is located nearer than the HII region at a distance of $9.4 \pm 0.3$ corresponding to the velocity of $-100$ km s$^{-1}$.  

G$338.3+0.0$: \cite{2016Supan} utilized HI absorption and $^{12}$CO emission spectra to find a distance of either $8.5$ or $13$ kpc. 
The tangent point velocity corresponding to the rotation curve of \cite{2014ReidRC} is $\sim120$ km s$^{-1}$.
The \cite{2016Supan} HI absorption spectra show absorption up to the tangent point. 
Thus, we resolve the kinematic distance ambiguity  and place the SNR at distance of $13 \pm 0.4$ corresponding to a velocity of $-31$ km s$^{-1}$.

The SNRs G$341.9-0.3$, G$342.0-0.2$ and G$343.1-0.7$ have no previously estimated distances. 
However, \cite{1998Koralesky} detected OH masers towards the remnants and we use their LSR velocity to estimate the distances. This places the SNRs G$341.9-0.3$ and G$342.0-0.2$ at a distance of $15.8 \pm 0.6$ kpc. 
For G$343.1-0.7$, $v_{\textrm{LSR}} = -70$ km s$^{-1}$ and which yields the near distance of $4.9 \pm 0.2$ kpc.

The distance to the SNR G$346.6 -0.2$ was presented by \cite{1998Koralesky} as either the near distance of 5.5 kpc or the far distance of 11 kpc corresponding to a radial velocity of $76.0$ km s$^{-1} $. 
\cite{2017Auchettl} used the tangent point distance of 8.3 kpc in their XMM-Newton study. 
We adopt the far distance of $10.4 \pm 0.2$ kpc consistent with \cite{2017Auchettl}.

G$348.5+0.0$: \cite{2012Tian} presented an upper limit to the distance to the remnant as 6.3 kpc. 
With HI absorption not present at $-107$ km s$^{-1}$, the corresponding distance  would be an upper limit (7.8 kpc with the rotation curve parameters presented by \cite{2014ReidRC}). 
The OH maser toward the SNR at $-100$ km s$^{-1}$ \citep{1998Koralesky} places the SNR is at a distance of $7.2 \pm 0.2$ kpc.

G$354.8-0.8$: The SNR has no previously known distance. We use the OH maser towards the SNR reported by \cite{1998Koralesky} at $v_{\textrm{LSR}} = -70$ km s$^{-1}$ to estimate the distance to the SNR. 
The kinematic distance ambiguity for the SNR cannot resolved and the distances are either $7.7 \pm 0.2$ or $8.9 \pm 0.2$ kpc. We take the average to place the SNR at a distance of  $8.3 \pm 0.8$ kpc.

\section{Radial Distribution and total number of Galactic SNRs} \label{Results}

\indent Detection of SNRs in the Galaxy is difficult because they are extended objects and the Galaxy has a diffuse synchrotron background and large number of HII regions which emit at the same radio frequencies as SNRs. Another reason is that SNRs can be similar or lower surface brightness than the background and confusing sources. Thus radio observations only detect some fraction of all SNRs in the Galaxy. This effect of detecting only a fraction of the SNRs is known as selection effects. We can account for selection effects in two ways: empirically defining correction factors which are the inverse of the selection function (which is less than 1); or as a functional form of the selection function. We refer to these two methods as Method 1 and Method 2.

\subsection{Method 1}

\begin{figure}
\centering
\includegraphics[width=\columnwidth]{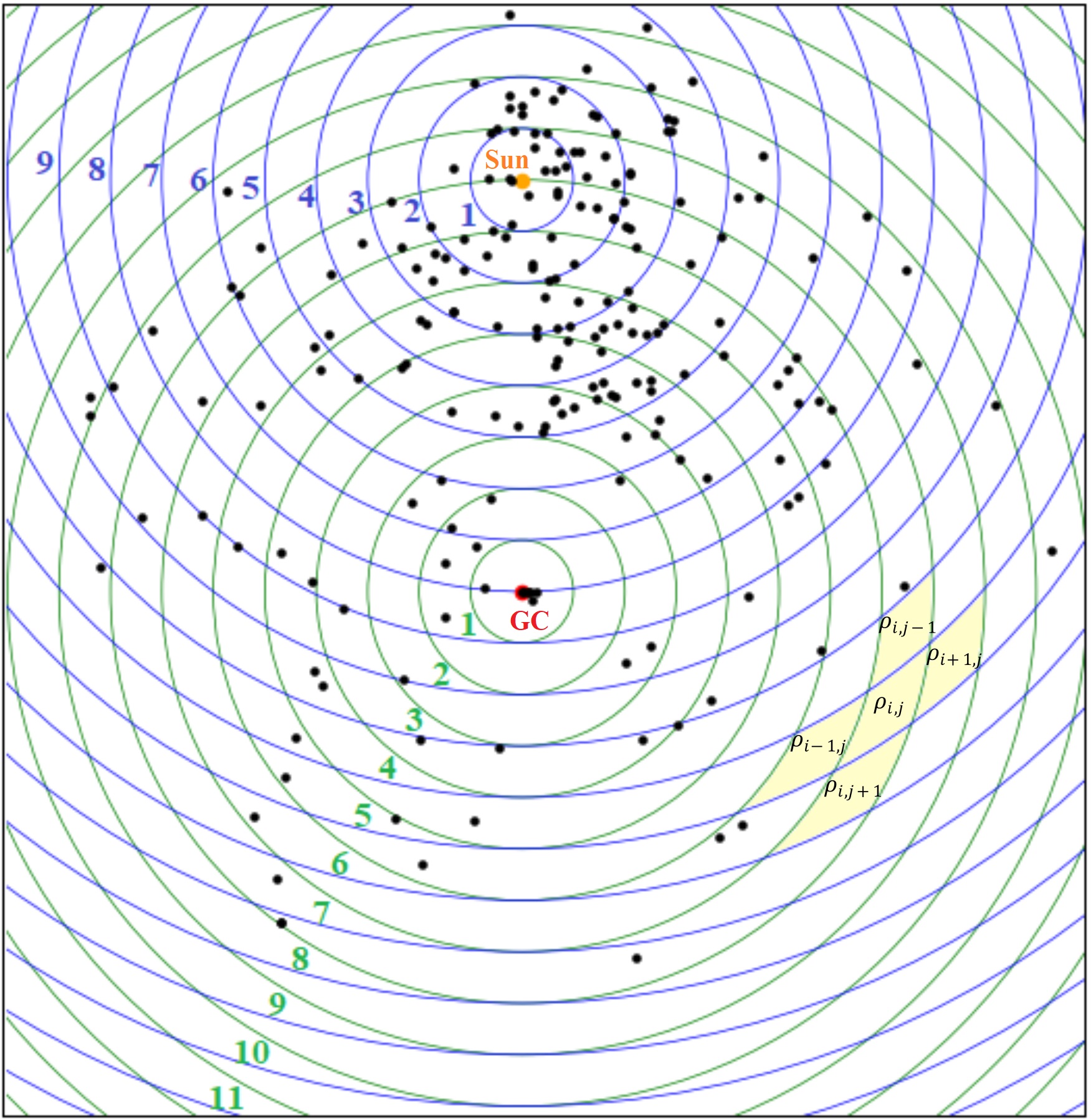}
\caption {Positions of Galactic SNRs from Table \ref{tab:1} projected onto the Galactic plane. 
The black dots show the positions of the SNRs. 
The concentric green rings ($i$: numbered in green) are at radii multiples of 1.0425 kpc centred at the Galactic center (GC; red circle) and the  blue rings ($j$: numbered in blue) are at radii multiples of 1.0425 kpc centred at the Sun (yellow circle). 
The yellow shaded regions show examples of the label system.}
\label{fig:1}
\end{figure}

\begin{figure}
\centering
\includegraphics[width=\columnwidth]{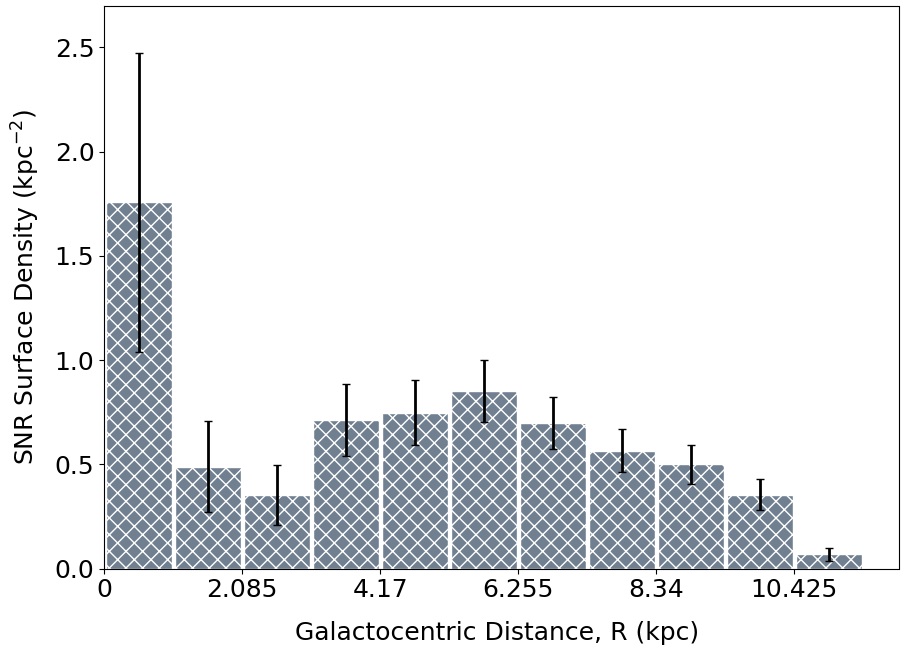}
\caption { The observed Galactic SNR densities (uncorrected).}
\label{fig:4}
\end{figure}

For Method 1, to determine the true surface densities we make the following assumptions:
\begin{enumerate}
\setlength\itemsep{1ex}
  \item The underlying surface density of SNRs is circularly symmetric about the GC;
  \item The underlying density is symmetric about line joining GC and Sun ($R_0$);
  \item The correction factors depend only on distance from the Sun ($d$).
\end{enumerate}

Due to observational selection effects, SNRs are concentrated near the Sun. 
Figure \ref{fig:1} shows the projection of the known Galactic SNRs with distances on the Galactic plane. 
The concentric circles around the GC and Sun (Figure \ref{fig:1} green and blue circles, respectively) were constructed with radii in multiples of $R_0/8$. 
The regions of interest are the smaller regions bound by the lines. We adopt the distance to the Sun ($R_0$) as $8.34$ kpc \citep{2014ReidRC}.\\

\subsubsection{Determination of Correction Factors and Surface Densities}

\indent To determine the the surface densities we first smooth the observed densities. 
Because the densities vary smoothly with $R$ and $d$, the 2-dimensional smoothing is done using 
\begin{equation} \label{eq1}
\rho'_{\textrm{\scaleto{i,j}{6pt}}} = (\rho_{\textrm{\scaleto{i,j-1}{6pt}}} + \rho_{\textrm{\scaleto{i-1,j}{6pt}}} + 4\rho_{\textrm{\scaleto{i,j}{6pt}}} + \rho_{\textrm{\scaleto{i,j+1}{6pt}}} + \rho_{\textrm{\scaleto{i+1,j}{6pt}}})/8.
\end{equation}
Here $\rho'$ is the smoothed surface density and $i$ \& $j$ (green and blue numbered rings in Figure \ref{fig:1}) correspond to the circles from the GC ($R$) and Sun ($d$), respectively \citep{1989LeahyXinji}. 
An example of the labelling is shown in Figure \ref{fig:1} (yellow shaded regions). \\
\indent However, equation (\ref{eq1}) is valid only for the regions that exclude the regions intersecting the extended Sun-GC line. 
There are 6 regions surrounding each region for the regions intersecting the extended Sun-GC line except for the regions where the Sun and GC are enclosed by the first ring ($i = 1$ and $j =1$, respectively). 
For the regions with 6 surrounding regions, we smooth the densities with the relation
\begin{equation}\begin{split}\label{eq2}
\rho'_{\textrm{\scaleto{i,j}{6pt}}} = (\rho_{\textrm{\scaleto{i-1,j+1}{6pt}}} + \rho_{\textrm{\scaleto{L:i,j+1}{6pt}}} + \rho_{\textrm{\scaleto{R:i,j+1}{6pt}}} + 6\rho_{\textrm{\scaleto{i,j}{6pt}}} + \\ \rho_{\textrm{\scaleto{L:i+1,j}{6pt}}} +  \rho_{\textrm{\scaleto{R:i+1,j}{6pt}}} + \rho_{\textrm{\scaleto{i+1,j-1}{6pt}}})/12.
\end{split}\end{equation}
The regions, excluding those intersecting the extended Sun-GC line, have mirror counterparts with the same $i$ \& $j$ indices. 
Therefore, in equation (\ref{eq2}) we denote the left and right regions with $L$ and $R$ subscripts.\\
\indent Finally, the four regions where the Sun and GC enclosed by the first ring (the regions: $\rho_{1,8}$, $\rho_{1,9}$, $\rho_{8,1}$ \& $\rho_{9,1}$) have only four regions surrounding them. 
The smoothing of these regions is done similar to the other regions with equation (\ref{eq1}).\\
\indent Next, the correction factors for the smaller regions are calculated. 
The correction factor ($C$) for a region in a particular ring is the density of the region in the line between the Sun and GC divided by the average density of the region of interest and its mirror counterpart. 
For an example, $C_{\scaleto{4,7}{6pt}}  =  \rho'_{\scaleto{4,5}{6pt}}/(\rho'_{\scaleto{L:4,7}{6pt}}/2 + R\rho'_{\scaleto{R:4,7}{6pt}}/2)$. 
This yields the correction factors for all regions except the regions in the line connecting the sun and the GC. \\
\indent For the $i = 1$ ring, we simply take the correction factor for $\rho_{1,8}$ region to be $\rho_{1,8}/\rho_{1,9}$. Furthermore, we assume no correction is factor is needed for the $j = 1$ ring. 
The assumption here is within $\sim1$ kpc radius of the Sun, observations are not affected by the selection bias. 
The rest of the correction factors (regions intersecting the sun-GC line) are taken to be the same as for the region at same $d$ but next larger $R$ (if $R < R_0$) or next smaller $R$ (if $R > R_0$). 
Because the regions in a ring are not equal to each other, the weighted average surface density of a each ring is calculated. \\
\indent To estimate the error in each ring, we take the standard error (SE) using the corrected surface densities of all regions in that ring. 
The standard error is calculated using $SE = \sigma_{s}/\sqrt{n}$, where $\sigma_s$ is the sample standard deviation and $n$ the number of regions. 
For the $i = 1$ ring, the uncorrected densities were used for the error estimate.

\subsubsection{Surface Density and Estimate of the Total Number of SNRs}

\begin{deluxetable}{lcc}[h]
\tablenum{2}
\label{tab:2}
\tablecaption{SNR Surface Density Estimates from Method 1}
\tablewidth{0pt}
\tablehead{
\colhead{R$^a$}       & \colhead{SNR Densities} &  \colhead{\# of SNRs}   \\
\colhead{(kpc)}       & \colhead{(kpc$^{-2}$)}  &  \colhead{}            
}
\startdata
$0 \qquad-	1.0425	$	&	$	2.12	\pm	0.78	$	&	$	7	\pm	3	$	\\
$1.0425	-	2.085	$   &	$	0.79	\pm	0.32	$	&	$	8	\pm	3	$	\\
$2.085	-	3.1275	$	&	$	0.97	\pm	0.14	$	&	$	16	\pm	2	$	\\
$3.1275	-	4.17	$	&	$	2.34	\pm	0.41	$	&	$	56	\pm	10	$	\\
$4.17	-	5.2125	$	&	$	2.35	\pm	0.29	$	&	$	72	\pm	9	$	\\
$5.2125	-	6.255	$	&	$	3.02	\pm	0.37	$	&	$	113	\pm	14	$	\\
$6.255	-	7.2975	$	&	$	2.96	\pm	0.26	$	&	$	131	\pm	11	$	\\
$7.2975	-	8.34	$	&	$	5.52	\pm	0.61	$	&	$	283	\pm	31	$	\\
$8.34	-	9.3825	$	&	$	3.44	\pm	0.48	$	&	$	200	\pm	28	$	\\
$9.3825	-	10.425	$	&	$	1.91	\pm	0.38	$	&	$	124	\pm	25	$	\\
$10.425	-	11.4675 $	&	$	0.56	\pm	0.12	$	&	$	40	\pm	9	$	\\
\hline
	Total		      &			        	& $1050 \pm 145 $ \\
\enddata
\tablecomments{a: The galactocentric distances $R$ are multiples of $R_0/8 = 1.0425$ kpc.}
\end{deluxetable}

\indent The densities and  estimated numbers of SNRs from Method 1 are given in Table \ref{tab:2}. 
From this work, there are now 215 SNRs with known distances (Table \ref{tab:1}). 
We have excluded SNRs with distances that are given only as a lower or upper limit. 
There are no SNRs between 12.51 and 15.6375 kpc. 
All SNRs in this sample except three are located at distances of less than $\sim11.5$ kpc from the GC. 
For the three exception, G$306.3-0.9$ is located at a galactocentric distance (R) of 16.5 kpc, G$184.6-5.8$ is at 11.7 kpc and G$141.2+5.0$ is at 11.7 kpc. 
G$306.3-0.9$ is the only SNR in its corresponding ring, so we exclude that ring from this analysis (because we do not expect to obtain a reasonable surface density, i.e. $\rho \sim 0$ kpc$^{-2}$). 
For the same reason, G$141.2+5.0$ and G$184.6-5.8$ were excluded from this analysis leaving 206 SNRs in the sample. \\
\indent The SNR surface densities estimated in Method 1 only extend up to a galactocentric distance $R \simeq 11.5$ kpc. 
In reality the surface densities are expected to extend to a considerably further indicating an incomplete sample. Therefore, the Method 1 estimated number of SNRs of $1050 \pm 145 $ is only a lower limit. 
Figure \ref{fig:5} shows the corrected observed distribution of SNR surface densities (top panel) and the number of SNRs at each galactocentric distance $R$ (bottom panel). \\

\subsection{Method 2} \label{method2}

\indent To determine the true distribution of SNRs in our Galaxy, we assume  circular symmetry of true SNR density about Galactic center ($R$) and that the selection function only depend on distance from the Sun ($d$) and Galactic longitude ($l$). In this case we use a functional form ($f_{\scaleto{sel}{5pt}}(d,l)$) that represents the selection effects.\\

\subsubsection{Fitting the Galactic SNR Distribution with Functional Forms}

\indent  The first step is to bin the SNRs in Galactic x and y, with each bin on a square grid on the Galactic plane centred on the Galactic center and $1.0425 \times 1.0425$ kpc ($R_0/8$) in size (see Figure \ref{fig:6} for the grid). 
The total number of bins is 400 ($20 \times 20$). 
Then the functional forms for the surface density, $\rho_{\scaleto{SNR}{3pt}}(R)$ and for the correction function $f_{sel}(d,l)$ were chosen. 
For the SNR surface density, $\rho_{\scaleto{SNR}{3pt}}(R)$ we select from the following set:
\begin{enumerate}
\setlength\itemsep{1ex}
  \item Exponential distribution (Exp) with scale length $H_\textrm{r}$,
  \begin{equation*}
	\rho_{\scaleto{SNR}{3pt}}(R) =  exp \left(- \frac{R}{H_\textrm{r}} \right).
  \end{equation*} 
  \item Power-law distribution with core (PL${_\textrm{C}}$) where core radius $R_c$ and $\alpha_1$ are fitting parameters,
  \begin{equation*}
	\rho_{\scaleto{SNR}{3pt}}(R) =  \left({1+ \frac{R}{R_c}} \right)^{-\alpha_1}.
  \end{equation*}  
  \item Gaussian distribution (GD) with a mean $\mu$ and standard deviation $\sigma$,
   \begin{equation*}
	\rho_{\scaleto{SNR}{3pt}}(R) =  exp \left(\frac{-(R - \mu)^2}{2\sigma^{2}} \right).
  \end{equation*} 
  \item Modified gamma function (MGF; \citealt{1977Stecker, 1998CaseBhat, 2021Verberne}),	  
    \begin{equation*}
	\rho_{\scaleto{SNR}{3pt}}(R) =  \left(\frac{R}{R_0} \right)^{\alpha_1} exp \left(-\beta \frac{R - R_0}{R_0} \right).
  \end{equation*}
  \item Distribution with a non-zero surface density at the GC (\citealt{1998CaseBhat} equation 15, hereafter referred to as CB98e15),  
  \begin{equation*}
	\rho_{\scaleto{SNR}{3pt}}(R) = \sin \left( \frac{\pi R}{R_a} + \theta_0 \right)e^{-\beta R}.
  \end{equation*}
  \item \cite{1963Sersic} profile,
  \begin{equation*}
	\rho_{\scaleto{SNR}{3pt}}(R) =  exp \left(-b_n \left[\left(\frac{R}{R_e} \right)^{1/n}  -1 \right] \right). 
  \end{equation*} 
\end{enumerate}
For Sersic function, $n$ and the half-light radius, $R_e$ are free parameters. 
There are a few approximations for the constant $b_n$ (see \cite{2005Graham}) in terms of the parameter $n$ for $n > 0.36$ \citep{1999Ciotti}. 
However, because the range of $n$ is unknown, we 
obtain $b_n$ by solving the equation $\Gamma(2n)  = 2\gamma(2n,b_n)$. 
Here $\Gamma$ is the (complete) gamma function and  $\gamma$ is the lower incomplete gamma function \citep{1991Ciotti}. \\
\indent For the correction function $f_{sel}(d,l)$, one of the following forms were chosen:
\begin{enumerate}
\setlength\itemsep{1ex}
  \item Exponential distribution (Exp) with scale length $H_\textrm{s}$,
  \begin{equation*}
	f_{\scaleto{sel}{5pt}}(d,l) =  exp \left(- \frac{d}{H_\textrm{s}} \right) \left(1 + B\cos(l - l_0) \right).
  \end{equation*} 
  \item Power-law distribution with core (PL${_\textrm{C}}$),
  \begin{equation*}
	f_{\scaleto{sel}{5pt}}(d,l) =  \left({1+ \frac{d}{R_s}} \right)^{-\alpha_2} \left( 1 + B\cos(l - l_0) \right),
  \end{equation*}   
\end{enumerate}
with $R_s$, $l_0$, $B$ and $\alpha_2$ fitting parameters. There are two main contributing factors to the selection effects: 1) Difficulty in detection of SNRs where the Galactic synchrotron background is bright, 2) The sensitivity of observations/surveys  in various regions of the Galactic plane. Therefore, a simple function was chosen to represent the Galactic synchrotron and sparseness of the data, a two-term Fourier series $(1 + B\cos(l - l_0)  )$. A simpler function (i.e. just the constant) does not describe the data well (see section \ref{EstimatedMethod2}). \\
\indent The model is the product of  $\rho_{\scaleto{SNR}{3pt}}(R)$ and $f_{\scaleto{sel}{5pt}}(d,l)$. 
This is fit to the observed distribution using $\chi^2$ minimization.
The $\chi^2$ function is given by
\begin{equation*}
\chi^2 = \sum_{i=1}^{N} \frac{(N_i - n_i)^2}{\sigma_i^2}. 
\end{equation*}
Here $N_\textrm{i}$ is the observed number of SNRs in each bin, $n_\textrm{i}$ is the model number of counts and $\sigma_i$ is the error of the observed counts. \\
\indent Because a significant number of bins are empty ($N_i = 0$) or have a small number of counts ($N_i \lesssim 10$), we use  the Y$^2$ statistics as described by \cite{2000Lucy} where  
\begin{equation*}
Y^2 = \nu + \sqrt{\frac{2\nu}{2\nu + \sum_{i} n_i^{-1}}} \left( \chi^2 -\nu \right) 
\end{equation*}
and $\nu$ is the number of degrees of freedom \citep{2007Press}. 
The errors for all the fitting parameters were found by finding the minimized Y$^2$ while keeping all parameters constant except for the parameter of interest. 

\subsubsection{SNR Surface Densities and Correction Functions from Method 2}

    The model parameters for the observed SNR densities were found by performing a $Y^2$ statistic minimization 
and are given in Table \ref{tab:3}. The parameter errors were calculated by fixing the minimized Y$^2$ at $+1\sigma$ and by calculating individual errors. Therefore, the true parameter errors are likely larger than the $1\sigma$ error.\\
\indent All models give good results except for case of the SNR density is a power-law distribution with $R_\textrm{c}$ and $\alpha_1$ left as free parameters. 
Here the parameters, $R_\textrm{c}$ and $\alpha_1$ are degenerate and provide equally good $Y^2$ values for large parameter values. 
Thus, we set the power-law exponent $\alpha_1 = 2$ and 1.5 in two cases, which yield the Plummer sphere and the Kuzman disk, respectively, \citep{2007SparkeGall}. \\
\indent For the SNR surface density as a Gaussian distribution, we left the mean as a free parameter or fixed  at $R = 0$ kpc. 
Both distributions yield a low $Y^2$ value and are included in Table \ref{tab:3}.\\ 
\indent Each model (product of surface density and selection function) has been superimposed on the observed SNR surface densities in Figure \ref{fig:6}. 
The differences in $Y^2$ statistic between all models shown are $\lesssim 2\sigma$. 
Therefore, we do not favour one model over another simply on the basis of the $Y^2$ statistic. 

\section{Discussion} \label{Discussion}

\subsection{Distance Estimates}

We compiled 215 SNRs with distances. 
Out of these we find distances to 6 SNRs where distances were previously unknown. 
A distance of $6.6 \pm 1.7$ kpc was determined for the SNR G$51.21 + 0.11$ using HI absorption, and the distances to the other five SNRs (G$5.5 + 0.3$, G$341.9-0.3$, G$342.0-0.2$, G$343.1-0.7$  and  G$354.8-0.8$) were determined by MC/maser associations. \\
\indent Distances to 135 of the SNRs are unchanged from the most recent literature values. These distances were obtained using HI absorption spectra, HI data analysis, maser and MC associations along with the \cite{2014ReidRC} rotation curve or by other methods (pulsar associations, parallax measurements etc.), where the radial velocity of the SNRs were unknown. \\
\indent  We have revised the distances or added new errors to 74 SNRs. 
The 74 SNRs include 9 distances that were significantly revised (see Section \ref{Galdist}). 
The distances for 20 of the 73 SNRs were improved at least by $10\%$. 
For 7 SNRs, we find the distances to be consistent with the literature values but present new estimates for errors. 
The distances to 38 SNRs slightly differ from the literature values due to the use of different rotation curves. 
Out of the the 38 SNRs, 15 did not have previously estimated errors. 
The 5 SNRs located near the GC do not have previously estimated errors (Table \ref{tab:1}). 
Due to the somewhat uncertain nature of their locations, we do not consider the errors to be accurate. 
For their errors, we utilize the error in $R_\textrm{0}$ of $\pm 0.16$ kpc presented by \cite{2014ReidRC}. \\
\indent \cite{2014Philstrom} places the SNR G$1.4 - 0.1$ near the GC. 
While we included this SNR in our sample given in Table \ref{tab:1}, we exclude from further analysis due to lack of strong evidence for its distance.  

\subsection{The Estimated Surface Density Using Method 1}

\begin{figure}
\centering
\includegraphics[width= \columnwidth]{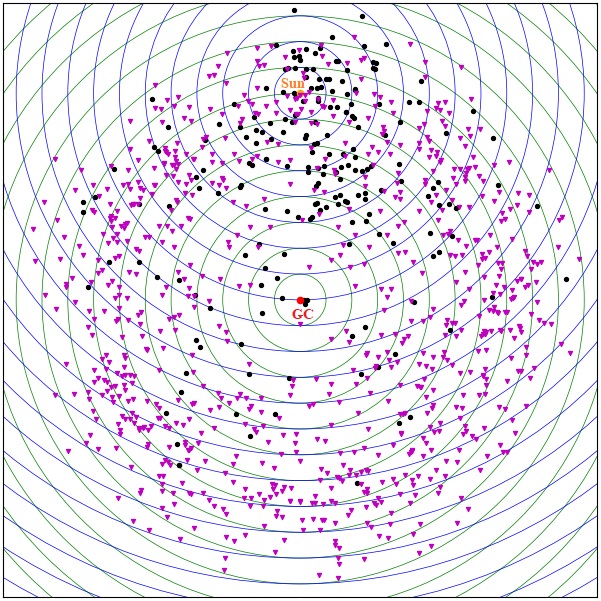}
\caption {The black circles are the positions of observed SNRs and the violet triangles are positions a simulated set of 1050 SNRs (Method 1). The ring radii are in multiples of 1.0425 kpc.}
\label{fig:3}
\end{figure}

\begin{figure}
\centering
\includegraphics[width=\columnwidth]{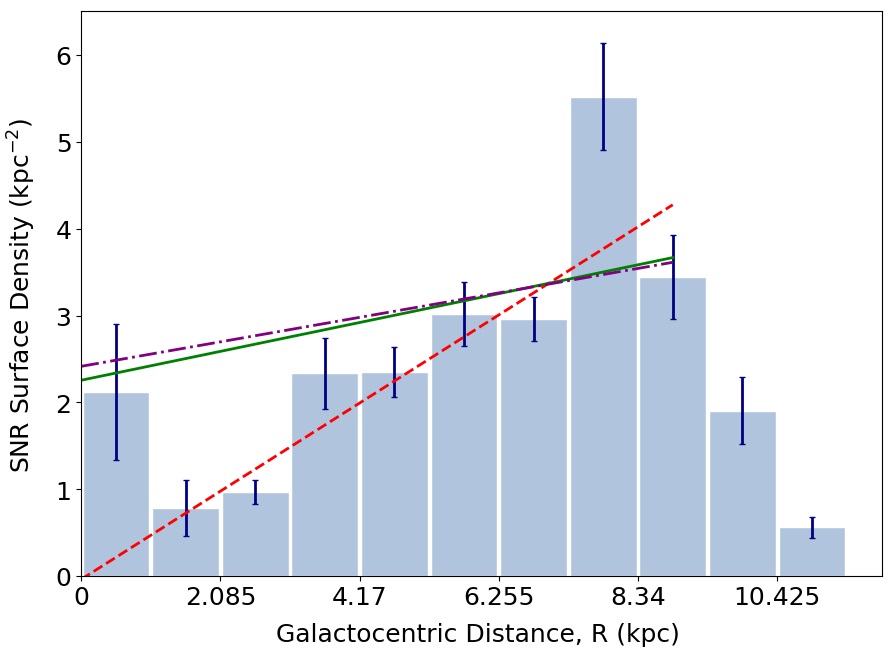}
\includegraphics[width=\columnwidth]{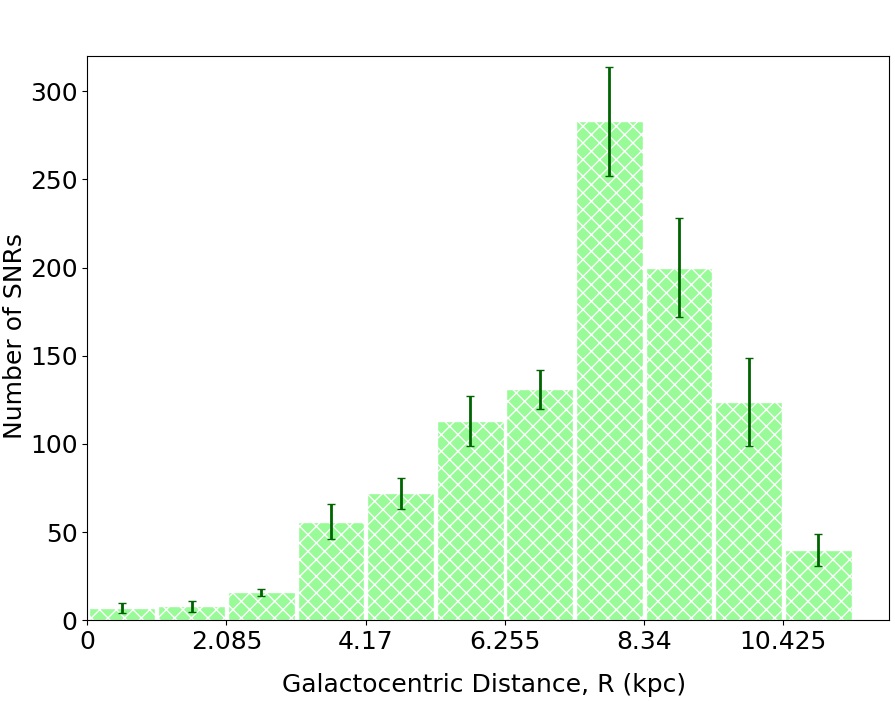}
\caption {Top: the surface density distribution estimated from Method 1. 
The dashed red line is best-fit line to SNR surface densities $< 9$kpc, the green solid line is the best-fit line omitting only the surface densities between 1.0425 and 3.1275 kpc and the violet dash-dot line omits the first three bins. 
Bottom: the distribution of the predicted total number of SNRs in each ring.}
\label{fig:5}
\end{figure}

With the SNR surface densities, total number of SNRs in the Galaxy ($R \lesssim 11.5$ kpc) can be estimated. 
Figure \ref{fig:3} shows the predicted number of SNRs (from Method 1) placed randomly  in each Galactic ring. 
There is a low density of SNRs for $ R \lesssim 3$ kpc. \\
\indent Figure \ref{fig:5} top panel shows SNR surface density distribution from Method 1. 
The red dashed line is the best-fit straight line for all bins with $ R < 9$ kpc. 
Figure \ref{fig:4} shows the uncorrected densities for comparison. \\
\indent The best-fit line when excluding the low densities in bins $1.0425 < R < 3.1275$ kpc is green solid line in Figure \ref{fig:5}. 
The best-fit line excluding the first 3 bins, (violet dot-dash line) is nearly the same. 
Both shows interpolated densities for the excluded region of $1.0425 < R < 3.1275$ kpc to be 2-3 times as large as the derived
model densities. The latter two linear fits predicts an additional $\sim 25 $ SNRs in this region. 
\indent A low density in the $1< R < 3$ kpc  region might be expected. 
The majority (about 85\%; \citealt{1994Tammann}) of the SNe are core-collapse (CC), where the the progenitor is a young massive star ($> 8M_{\sun}$).  
The deficit of observed star forming regions or high mass X-ray binaries at Galactocentric distance,  $R \lesssim 3$ kpc  \citep{2003Russeil, 2012Bodaghee} is consistent with the low density of  SNRs in this region. 
The Method 1 SNR distribution seem to be consistent with the distribution of star forming  complexes presented by \citep{2003Russeil}. 
Similar to the these complexes, the observed SNR distribution does not show any large scale spiral structure. 
The low density of HII regions at a Galactocentric radius, $R < 3.5 $ kpc is evident from the observations made by \cite{1981Lockman}, \cite{2009Anderson} and \cite{2012Anderson}. 
Additionally, the low density of Galactic young objects and masers at $R \lesssim 3$ was apparent in the study done by \cite{2020Shen} on the Galactic structure.\\
\indent Even though the majority of SNe are CC, the contribution of the SNRs resulted by type Ia explosions is significant.
 The distribution of type Ia SNRs in the Galaxy would not follow the distribution of star forming regions.  
It is possible that the majority of SNRs in the region  $R \lesssim 3$ kpc are type Ia. 
In fact, of the 215 SNRs in our sample, 16 SNRs lie in the region  $R \lesssim 3$ kpc and SN types of 8 of these are known. 
3 SNRs of the 8 are CC and 5 are type Ia.  
While the sample here is too small to form a definite conclusion, the observations are consistent with the deficit of CC SNRs in the region  $R \lesssim 3$ kpc. \\
\indent As a test for the Method 1 validity, we have conducted simulations to estimate the SNR density distribution.  For a chosen initial number of SNRs, we placed them in each ring ($i^{th}$ ring; Figure \ref{fig:1}) consistent with a surface density distribution that is combined with a correction function (see Section \ref{method2}). From the resultant distributions (simulated observed SNR distribution), we estimated the surface density distributions using Method 1. These tests have shown that the smoothing and corrections produce a total number of SNRs that is less than the initial input ($\sim 1/3.5$). The Method 1 SNR surface density estimations are sensitive to slight changes and the sparseness of the data leads to an underestimation of the total number of Galactic SNRs. Therefore, the total number of Galactic SNRs ($1050$) estimated using Method 1 is likely a lower limit and the total number of SNRs is estimated at roughly as $\sim3500$. 

\subsection{Estimated Surface Densities Using Method 2} \label{EstimatedMethod2}

The models and best-fit parameters for the surface density  and correction functions are given in Table \ref{tab:3}. 
For all models shown\footnote{We ran a number of other models of similar other functional forms with or same functional 
forms with fewer parameters- all of those models (not shown) had higher Y$^2$ by more than $2\sigma$.}, the 
Y$^2$ statistics show that the fits are equally good within $2\sigma$. 
The contour maps in Figure \ref{fig:6} are of the product of SNR surface density functions and correction functions ($A\rho_{\scaleto{SNR}{3pt}}(R)f_{\scaleto{sel}{5pt}}(d,l)$). 
The numbers of observed SNRs are shown by the green $1.0425 \times 1.0425$ kpc grid. 
From Figure \ref{fig:6}, it is seen that the model distributions appear to be in agreement with the observed distribution of SNRs. \\
\indent The exponential profiles for the SNR surface density yield reasonable Y$^2$ values and scale lengths ($R_\textrm{H}$). 
The  Galactic thin disk scale length given by \citep{2016BlandHawthorn} is $2.6 \pm 0.5$ kpc. 
If the scale length ($R_\textrm{H}$) is fixed in the exponential profiles (see models 1 and 2 in Table \ref{tab:3}), the resulting Y$^2$ values are $167.21$ and $166.83$ for the two selection functions, respectively. 
These Y$^2$ values are $< 1 \sigma$ and $< 1.3 \sigma$ higher than the  model 1 and 2 values (Table \ref{tab:3}), indicating that the scale lengths are consistent with the Galactic thin disk scale length. 
However, the SNR density functions significantly vary from each other, implying that the data 
is insufficient to differentiate between the listed models. \\
\indent Figure \ref{fig:7} shows the radial distributions of the GD, GD$_{\mu = 0}$, MGF, CB98e15 and S{\'e}rsic profiles.  
The MGF  densities peak at $R = 4.1$ kpc and $R = 4.3$ kpc for the two correction functions (Exp and PL${_\textrm{C}}$, respectively) and agree with the peak of the radial distribution (at $R \approx 5$ kpc) provided by \cite{1998CaseBhat} within error. \\
\indent The CB98e15 profile yields a non-zero density at the center and the parameters (see Table \ref{tab:3}) are consistent with the parameters estimated by \citealt{1998CaseBhat}. 
The density function is zero beyond  $R= R_a(1 - \theta_0/\pi)$  ($R = 12.3$ and $12.5$ kpc for exponential and power-law with core correction functional forms, respectively). \\
\indent The parameter $n$ that describes the shape of the S{\'e}rsic profile is low ($\sim0.25)$ compared to common values of $n = 4$ \& $1$ \citep{1999Ciotti}. 
However, the half-light radius estimated in the S{\'e}rsic profile ($R_\textrm{e})$ of $\sim6$ kpc is consistent with the expected value for a Milky way-like Galaxy \citep{2020Chamba}.  \\ 
\indent If we assume the observed SNRs for an area of $ d \lesssim 1$ kpc are complete near the Sun, we find the SNR surface density to be  $4.4 \pm 1.1 $ kpc$^{-2}$. 
The model densities yield  $5.1$ to $6.1$  kpc $^{-2}$ for the exponential correction function, whereas the power law  correction function yields higher densities of $7.2$ to $8.9$ kpc$^{-2}$. 
These are higher than the observed density near the sun, indicating incompleteness by factor $\sim$2 within 1 kpc of the sun (power law with core) or by factor $\sim$105 (exponential).  \\
\indent Apart from the models provided in Table \ref{tab:3}, we have investigated a few more complex models for the SNR surface density functional form. Among them, the sum of two Gaussian distributions and a distribution with a Galactic bar \citep{2015Wegg} were tested. 
However, we found that the data was inadequate to determine best-fit parameters (the parameters are degenerate).\\
\indent If the angular dependence is omitted, the models yield an increase in Y$^2$ value which is greater than $3\sigma$ and thus statistically significant, indicating the angular dependence of the correction function is necessary. 
We also tested the simplest model for the SNR surface density function: $\rho_{\scaleto{SNR}{3pt}}(R)$ is a constant. 
In this case, the parameters for the models ($A f_{\scaleto{sel}{5pt}}(d,l)$) are degenerate for both correction functions. 
The Y$^2$ values for both selection functions are $\sim172$  and $> 2\sigma$ higher compared to the exponential density distributions (models 1 \& 2). If the angular dependence is also ignored, the Y$^2$ values are higher by $> 3\sigma$ compared to models 1 and 2 in Table \ref{tab:3}. 

Therefore, with the current sample of SNRs, we conclude that more complex SNR surface density functions are not justified by the data, and that simpler functions for SNR surface density or correction function are not allowed by the current data.
   
\begin{figure}
\centering
\includegraphics[width=\columnwidth]{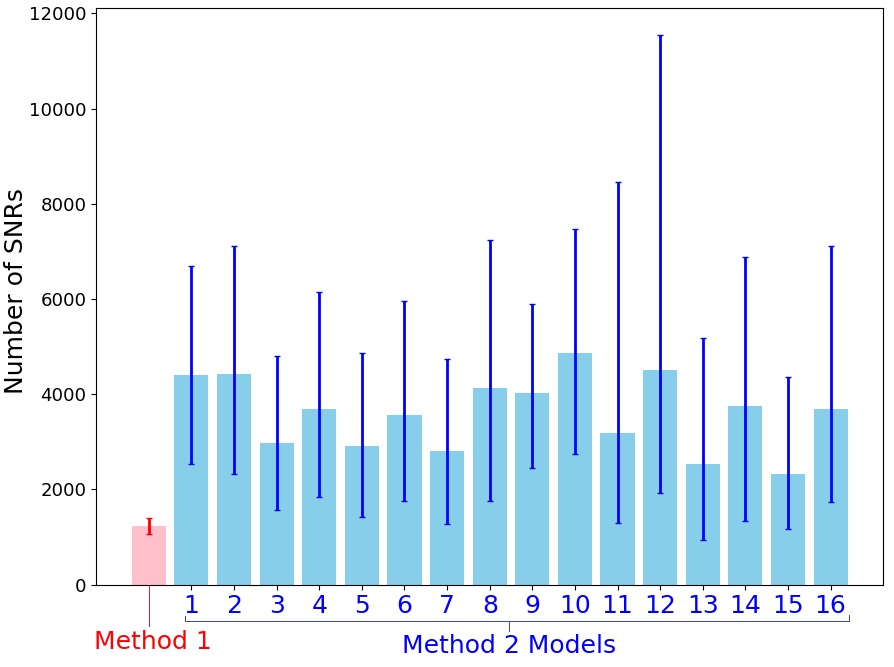}
\caption {Total number of SNRs estimated using Method 1 (pink bar \& red error) and Method 2 (light blue bars 
\& blue errors). The Method 2 numbers correspond to the model numbers in Table \ref{tab:3}. 
All numbers are calculated over the region $0<R \le 11.5 $ kpc.   }
\label{fig:8}
\end{figure}

\begin{figure*}
\centering
\includegraphics[width=\textwidth]{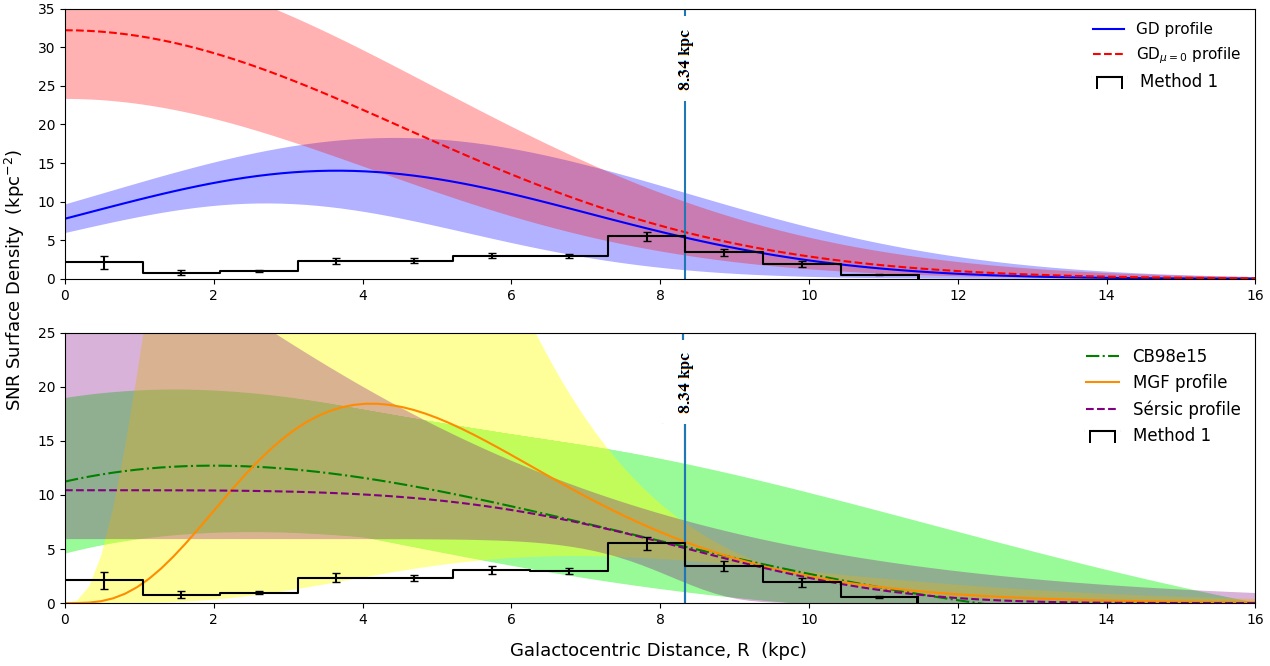}
\caption {SNR surface densities from Method 2. Top: GD: blue solid curve and GD$_{\mu = 0}$ profile: red dashed curve. Bottom: MGF: orange solid line, CB98e15: green dash-dot curve and S{\'e}rsic: purple dashed curve. 
All surface densities are for the case of  an exponential correction function. 
The shaded regions are the lower and upper limits of the similarly-coloured profiles.
}
\label{fig:7}
\end{figure*}

\subsection{Comparison Between Method 1 and Method 2 Galactic SNR Surface Densities}

\indent Figure \ref{fig:8} shows the comparison of Method 1 SNR total number (pink bar) and Method 2 total SNR numbers (light blue bars), both integrated over the area of the Galaxy with $R \le 11.5 $ kpc. 
The CB98e15 and S{\'e}rsic profiles in the case that the correction function is an exponential distribution  yield  lower limit numbers of 940 and 1160 SNRs, respectively. 
While these are consistent with Method 1 number within error, the Method 2 density functions do not agree with the Method 1 densities.
\\
\indent Comparisons of the Method 2 density functions with each other and the Method 1 densities are shown in Figure \ref{fig:7}.
The Method 2 densities are consistent within errors (the error bands overlap) outside $R \sim 7$ kpc. 
The GD$_{\mu=0}$ density is inconsistent (higher than) the GD inside $R \sim 3.5$ kpc and higher than the other densities inside other $R$ values, with $R$ value dependent on the function.
The exponential surface density (slightly higher than even the GD$_{\mu=0}$ density) is consistent with the Method 1 density 
only for a $R \ga 8$ kpc and is similarly (like the GD$_{\mu=0}$ density) inconsistent with the other Method 2 densities in the inner Galaxy. 

\indent  In summary, from Figure \ref{fig:1}, we see that the observed number of SNRs is not complete and selection effects are clearly present. For Method 1, adding a few observed SNRs in a region where there are none, considerably increases (by $>\sim0.1$ kpc$^{-2}$) the smoothed and corrected densities. 
Figure~\ref{fig:7} provides a summary of the results: the current data results in density functions with large error bands,
thus we conclude that identification of more observed SNRs is needed to constrain better Galactic SNR densities.  

\subsection{The Total Number of Galactic SNRs}

While method 1 produced $\sim3500$ total number of SNRs, Method 2 produced $\sim2400$ to $\sim5600$ SNRs. With $\sim300$ observed Galactic SNRs (confirmed: \cite{2019GreenCat}), the majority of the SNRs remain to be discovered. \\
\indent Using the THOR + VGPS combined survey, \cite{2017Anderson} presented 76 SNR candidates covered by the region between Galactic longitudes $17.5\arcdeg$ and $67.4\arcdeg$ and latitudes $-1.25\arcdeg$ and $+1.25\arcdeg$. Of the 294 SNRs there are only 52 SNRs in the region covered by THOR. If the \cite{2017Anderson} SNR candidates are confirmed, it brings the total Galactic SNRs in the region to 128 SNRs which is a $\sim150\%$ increase.\\
\indent \cite{2019Hurley} presented 27 SNR candidates using the GLEAM survey covered over Galactic longitudes, $345\arcdeg < l < 60\arcdeg$ and $180\arcdeg < l < 240\arcdeg$. Here 25 candidates are located at $345\arcdeg < l < 60\arcdeg$ and 2 are located at $180\arcdeg < l < 240\arcdeg$. Which increases the total numbers of SNRs if confirmed, by $17\%$ and $22\%$ respectively, for the two regions.\\
\indent Considering both  \cite{2017Anderson} and  \cite{2019Hurley} SNR candidates, an increase of $150\%$ is a conservative estimate for $17.5\arcdeg < l < 67.4\arcdeg$ and extrapolating it to the Galaxy, if confirmed there may be as many as $\sim750$ SNRs in the Galaxy. However, the total number of SNRs still fall short of the predicted number of a few thousand. \\  
\indent  The majority of the SNRs are located in the Galactic plane. At least $90\%$ of the 215 Galactic SNRs where distances are known (Table \ref{tab:1}) are at a Galactic height, $z \la  200 $ pc. Most of the SNRs found in the Galactic plane are CC and can be associated with star forming regions. It is likely that many SNRs are not detected because of the confusion caused by objects (SNRs, star forming regions etc.) in the foreground. \\
\indent The detection of faint SNRs depends on the sensitivity of observations/surveys. With increased sensitivity it is likely objects that are not associated with each other could be easily detected to lessen the confusion. However, the brightness of the foreground objects dominates in any particular region of the Galaxy and detection of the background objects would be a difficult task. Furthermore, SNRs will be difficult to detect where the Galactic synchrotron background is bright, which is mainly in the Galactic plane. While the predicted number of SNRs for the Galaxy is a few thousand, because of the above reasons the observations/detections of new SNRs could be limited to a few hundred.

\section{conclusion} \label{conclusion}

In order to examine the distribution of Galactic SNRs, accurate distances are essential. Using HI absorption, we estimate the distance of $6.6 \pm 1.7$ kpc to the SNR  G$51.21 + 0.11$, where the distance was previously unknown. Furthermore, using maser/MC associations, distances to five more SNRs where the distance were previously unknown were estimated. While distances to 135 SNRs remained unchanged, distances to 22 SNRs were significantly revised ($>10\%$). We have produced a list of 215 Galactic SNRs with distances with enough accuracy to test models.  \\
\indent We utilized two methods in order to test density distributions for Galactic SNRs. Method 1 employs correction factors for the selection effects to derive a binned density. However, the density distribution estimated from this method is sensitive to slight changes. Because of the sparseness of the data and weaknesses of the model,  Method 1 yields only ~1000 SNRs, while tests show this actual number may be closer to $\sim3500$ Galactic SNRs  \\ 
\indent For the second method we test functional forms for the SNR surface density and the correction factor using fits to the observed data. 
Figure \ref{fig:7} shows the models are all consistent in the region near the sun but differ most near the Galactic center.
 The best-fit parameters for more complex models cannot be determined because of the limited sample size of observed SNRs.   \\
\indent Method 2 yields $\sim2400$ to $\sim5600$ SNRs.. 
The predictions for the total number of SNRs in the Galaxy are all $> 1000$ 
Additional observed SNRs are required to obtain results with small uncertainties for total number of SNRs and their radial distribution. \\

We thank Dr. B. Karchewski for helpful comments. This work was supported by a grant from the Natural Sciences and Engineering Research Council of Canada.

\startlongtable
\begin{deluxetable*}{llcccccc}
\tablenum{1}
\label{tab:1}
\tablecaption{Galactic SNRs with reliable distances }
\tablewidth{0pt}
\tablehead{
\colhead{\#} & \colhead{SNR$^a$}  & \colhead{Literature} &  \colhead{Method}  & \colhead{$V_{LSR} $} &  \colhead{Ref} & \colhead{Revised} & \colhead{Galactocentric}  \\
\colhead{} & \colhead{}  & \colhead{dist (kpc)}  &  \colhead{}  & \colhead{(km s$^{-1}$)} & \colhead{} & \colhead{dist$^b$ (kpc)} & \colhead{dist (kpc)} 
}
\startdata
1	&	       G$0.0+0.0$      	&	8	&	Galactic center	&	       \nodata 	&	1	&	       8.34 $\pm$ 0.16     	&	0	\\
2	&	       G$0.3+0.0$      	&	8.5	&	Galactic center	&	       \nodata 	&	2	&	       8.34 $\pm$ 0.16     	&	0.04	\\
3	&	       G$0.9+0.1$      	&	8.5	&	Galactic center	&	       \nodata 	&	3	&	       8.34 $\pm$ 0.16     	&	0.13	\\
4	&	       G$1.0-0.1$      	&	8	&	Galactic center	&	       \nodata 	&	1	&	       8.34 $\pm$ 0.16     	&	0.15	\\
5	&	       G$1.4-0.1$      	&	8.5	&	       \nodata 	&	       \nodata 	&	4	&	       \nodata                     	&	   \nodata           	\\
6	&	       G$1.9+0.3$      	&	8.5	&	Galactic center	&	       \nodata 	&	5	&	       8.34 $\pm$ 0.16     	&	0.28	\\
7	&	       G$4.5+6.8$      	&	       $5.1^{+0.8}_{-0.7}$     	&	Proper motion + $v_{Shock}$	&	       \nodata 	&	6	&	       \nodata                         	&	       $3.3^{+0.8}_{-0.7}$     	\\
8	&	       G$5.4-1.2$      	&	5.2	&	Kinematic distance	&	27	&	7,8	&	   5.0 $\pm$ 0.3             	&	       3.4 $\pm$ 0.3 	\\
9	&	       G$5.5+0.3$      	&	       \nodata                 	&	       \nodata                 	&	12.5	&	9	&	       3.2     $\pm$ 0.7           	&	   5.1     $\pm$ 0.7 	\\
10	&	 G$5.7-0.1^\ddagger$ 	&	 2.9 $\pm$ 0.3   	&	Optical extinction	&	 12.8$^*$  	&	10	&	 3.0 $\pm$ 0.7      	&	 5.3  $\pm$ 0.7 	\\
11	&	       G$6.4-0.1$      	&	       1.9     $\pm$ 0.3           	&	Kinematic distance	&	7	&	11	&	   1.8     $\pm$ 0.3           	&	   6.6     $\pm$ 0.3 	\\
12	&	 G$7.5-1.7^\ddagger$ 	&	1.7	&	PWN association	&	 \nodata   	&	12	&	 \nodata               	&	6.7	\\
13	&	       G$7.7-3.7$      	&	       4.5     $\pm$ 1.5           	&	Estimation (z distance)	&	   \nodata 	&	13	&	       \nodata             	&	   3.9     $\pm$ 1.4 	\\
14	&	       G$8.3+0.0$      	&	16.4	&	Kinematic distance	&	2.6	&	14	&	   15.9 $\pm$ 1        	&	   7.7     $\pm$ 1.0 	\\
15	&	       G$8.7-0.1$      	&	4.5	&	Kinematic distance	&	36	&	7	&	   4.5     $\pm$ 0.3           	&	   3.9     $\pm$ 0.3 	\\
16	&	       G$9.7+0.0$      	&	4.7	&	Kinematic distance	&	43	&	7	&	   4.8     $\pm$ 0.2           	&	   3.7     $\pm$ 0.2 	\\
17	&	       G$9.9-0.8$      	&	4	&	Kinematic distance	&	31	&	14	&	   3.8     $\pm$ 0.3       	&	       4.6     $\pm$ 0.3 	\\
18	&	       G$11.0+0.0$     	&	       2.4     $\pm$ 0.7           	&	Optical extinction	&	   \nodata 	&	10	&	       \nodata                 	&	       6       $\pm$ 0.7 	\\
19	&	       G$11.2-0.3$     	&	7.2	&	Kinematic distance	&	32	&	14	&	   3.7     $\pm$ 0.2           	&	   4.8     $\pm$ 0.3 	\\
20	&	       G$12.0-0.1$     	&	10	&	Pulsar association	&	       \nodata 	&	15	&	       \nodata                     	&	2.5	\\
21	&	       G$12.2+0.3$     	&	11.8	&	Kinematic distance	&	50	&	14	&	   11.6 $\pm$ 0.3      	&	   3.9     $\pm$ 0.3 	\\
22	&	       G$12.8+0.0$     	&	4.7	&	Pulsar association	&	       \nodata 	&	16	&	       \nodata                     	&	3.9	\\
23	&	       G$13.3-1.3$     	&	       3.3 $\pm$ 1.3       	&	Kinematic distance	&	30$^{**}$  	&	17	&	       3.1     $\pm$ 0.3           	&	   5.3     $\pm$ 0.3 	\\
24	&	       G$15.1-1.6$     	&	2.1	&	Optical extinction	&	   \nodata 	&	10	&	       \nodata                 	&	       6.3      	\\
25	&	       G$15.4+0.1$     	&	       9.3     $\pm$ 1.3           	&	Kinematic distance	&	95	&	18	&	   9.8     $\pm$ 1.3           	&	   2.8     $\pm$ 0.8 	\\
26	&	       G$15.9+0.2$     	&	       11.5 $\pm$ 4.5          	&	Kinematic distance	&	       \nodata 	&	19	&	       \nodata                 	&	       4.2     $\pm$ 3.8 	\\
27	&	       G$16.0-0.5$     	&	3.4	&	CO observations	&	       \nodata 	&	20	&	       \nodata                     	&	5.2	\\
28	&	       G$16.7+0.1$     	&	14	&	Kinematic distance	&	20	&	19	&	   13.9 $\pm$ 0.4      	&	   6.4     $\pm$ 0.4 	\\
29	&	       G$17.4-2.3$     	&	       $>2.3$                      	&	Estimation 	&	   \nodata 	&	21	&	       \nodata                         	&	       $>8$              	\\
30	&	 G$18.0-0.7^\ddagger     $ 	&	 3.1 $\pm$ 0.2       	&	Optical extinction	&	 \nodata       	&	10	&	   \nodata             	&	   5.5 $\pm$ 0.2 	\\
31	&	       G$18.1-0.1$     	&	       6.4     $\pm$ 0.2           	&	Kinematic distance	&	103.74	&	22	&	       \nodata                         	&	       3.0 $\pm$ 0.2 	\\
32	&	       G$18.6-0.2$     	&	       4.4     $\pm$ 0.2           	&	Kinematic distance	&	62.84	&	22	&	       \nodata             	&	   4.4     $\pm$ 0.2 	\\
33	&	 G$18.76-0.072^\dagger$ 	&	      4.7     $\pm$ 0.3       	&	Kinematic distance	&	70	&	23	&	   \nodata                     	&	   4.2 $\pm$ 0.2 	\\
34	&	       G$18.8+0.3$     	&	       13.8 $\pm$ 0.4      	&	Kinematic distance	&	21.35	&	22	&	       \nodata                 	&	       6.5     $\pm$ 0.4 	\\
35	&	       G$18.9-1.1$     	&	       2.1     $\pm$ 0.4           	&	Kinematic distance	&	23.27	&	24	&	       \nodata                         	&	       6.4     $\pm$ 0.4 	\\
36	&	       G$20.0-0.2$     	&	       11.2 $\pm$ 0.3      	&	Kinematic distance	&	66.4	&	22	&	       \nodata                 	&	       4.4     $\pm$ 0.2 	\\
37	&	       G$21.5-0.9$     	&	       4.4     $\pm$ 0.2           	&	Kinematic distance	&	67.79	&	22	&	       \nodata                 	&	       4.5     $\pm$ 0.2 	\\
38	&	 G$21.8-3.0^\dagger$        	&	  0.33 $\pm$      0.1     	&	Kinematic distance	&	5.9	&	25	&	  \nodata                        	&	   8.0 $\pm$ 0.2 	\\
39	&	       G$21.8-0.6$     	&	       5.6     $\pm$ 0.2           	&	Kinematic distance	&	93.35	&	22	&	       \nodata                 	&	       3.8     $\pm$ 0.2 	\\
40	&	       G$22.7-0.2$     	&	       4.7     $\pm$ 0.2           	&	Kinematic distance	&	76.63	&	22	&	       \nodata                 	&	       4.4     $\pm$ 0.2 	\\
41	&	       G$23.3-0.3$     	&	       4.8     $\pm$ 0.2       	&	Kinematic distance	&	78.51	&	22	&	       \nodata                 	&	       4.4     $\pm$ 0.2 	\\
42	&	       G$24.7-0.6$     	&	       3.8     $\pm$ 0.2           	&	Kinematic distance	&	60.67	&	26	&	       \nodata                 	&	       5.1     $\pm$ 0.2 	\\
43	&	       G$24.7+0.6$     	&	       3.5     $\pm$ 0.2           	&	Kinematic distance	&	54.6	&	27	&	       \nodata                         	&	       5.4     $\pm$ 0.2 	\\
44	&	       G$25.1-2.3$     	&	2.9	&	Kinematic distance	&	39.2	&	28	&	       2.7     $\pm$ 0.3           	&	   6       $\pm$ 0.3 	\\
45	&	 G$26.6-0.1^\ddagger$	&	1.3	&	Absorption column	&	 \nodata   	&	29	&	 \nodata                       	&	7.2	\\
46	&	       G$27.4+0.0$     	&	       5.8     $\pm$ 0.3       	&	Kinematic distance	&	99.95	&	22	&	       \nodata         	&	       4.2     $\pm$ 0.2 	\\
47	&	       G$28.6-0.1$     	&	       9.6     $\pm$ 0.3       	&	Kinematic distance	&	85	&	26	&	   \nodata         	&	       4.6     $\pm$ 0.1 	\\
48	&	       G$28.8+1.5$     	&	       3.4     $\pm$ 0.6       	&	Optical extinction	&	       \nodata 	&	10	&	       \nodata         	&	       5.6     $\pm$ 0.4 	\\
49	&	 G$29.4+0.1^\ddagger$	&	 6.5 $\pm$ 1     	&	Kinematic distance	&	97	&	30	&	   \nodata         	&	 4.2   $\pm$ 0.2 	\\
50	&	       G$29.6+0.1$     	&	       4.7     $\pm$ 0.3       	&	Kinematic distance	&	80.16	&	27	&	       \nodata         	&	       4.8     $\pm$ 0.2 	\\
51	&	       G$29.7-0.3$     	&	       5.6     $\pm$ 0.3       	&	Kinematic distance	&	95	&	22	&	   \nodata         	&	       4.4     $\pm$ 0.2 	\\
52	&	 G$31.299-0.493^\dagger$	&	      5   $\pm$ 0.3   	&	Kinematic distance	&	85	&	23	&	     \nodata         	&	 4.8 $\pm$     0.2  	\\
53	&	       G$31.9+0.0$     	&	       7.1     $\pm$ 0.4       	&	Kinematic distance	&	       $v_{\scaleto{TP}{3pt}}$	&	31	&	        \nodata     	&	   4.4     $\pm$ 0.1 	\\
54	&	       G$32.1-0.9$     	&	2	&	Optical extinction	&	   \nodata 	&	10	&	       \nodata         	&	6.7	\\
55	&	       G$32.4+0.1$     	&	11.8	&	Kinematic distance	&	43	&	14	&	   11.3 $\pm$ 0.3  	&	       6.2     $\pm$ 0.2 	\\
56	&	       G$32.8-0.1$     	&	       4.8     $\pm$ 0.3       	&	Kinematic distance	&	81.81	&	22	&	       \nodata         	&	       5       $\pm$ 0.2 	\\
57	&	       G$33.6+0.1$     	&	       3.5     $\pm$ 0.3       	&	Kinematic distance	&	57.9	&	22	&	       \nodata         	&	       5.8     $\pm$ 0.2 	\\
58	&	       G$34.7-0.4$     	&	       3       $\pm$ 0.3       	&	Kinematic distance	&	50.48	&	22	&	       \nodata         	&	       6.1     $\pm$ 0.2 	\\
59	&	       G$35.6-0.4$     	&	       3.8     $\pm$ 0.3       	&	Kinematic distance	&	63.67	&	22	&	       \nodata         	&	       5.7     $\pm$ 0.2 	\\
60	&	       G$38.7-1.3$     	&	4	&	Absorption column	&	   \nodata 	&	32	&	       \nodata         	&	5.8	\\
61	&	       G$39.2-0.3$     	&	       8.5     $\pm$ 0.5       	&	Kinematic distance	&	69.39	&	22	&	       \nodata         	&	       5.7     $\pm$ 0.2 	\\
62	&	       G$39.7-2.0$     	&	       4.9     $\pm$ 0.4       	&	Kinematic distance	&	77	&	33	&	   5.1     $\pm$ 0.4       	&	       5.5     $\pm$ 0.2 	\\
63	&	       G$40.5-0.5$     	&	3.4	&	Kinematic distance	&	55	&	34	&	       3.4     $\pm$ 0.4       	&	       6.1     $\pm$ 0.2 	\\
64	&	       G$41.1-0.3$     	&	       8.5     $\pm$ 0.5       	&	Kinematic distance	&	       \nodata 	&	22	&	       \nodata         	&	       5.9     $\pm$ 0.2 	\\
65	&	       G$41.5+0.4$     	&	       4.1     $\pm$ 0.5       	&	Kinematic distance	&	63.67	&	26	&	       \nodata         	&	       5.9     $\pm$ 0.2 	\\
66	&	       G$42.8+0.6$     	&	7.7	&	Pulsar association	&	       \nodata 	&	35	&	       \nodata         	&	5.9	\\
67	&	       G$43.3-0.2$     	&	       11.3$\pm$ 0.4   	&	Kinematic distance	&	12.55	&	22	&	       \nodata         	&	       7.8     $\pm$ 0.3 	\\
68	&	       G$43.9+1.6$     	&	       3.1     $\pm$ 1.2       	&	Kinematic distance	&	50	&	36	&	   \nodata         	&	       6.4     $\pm$ 0.5 	\\
69	&	 G$44.5-0.2^\ddagger$	&	4.1	&	Kinematic distance	&	60	&	37	&	 4.1 $\pm$ 0.7      	&	 6.1  $\pm$ 0.3 	\\
70	&	       G$46.8-0.3$     	&	       8.4     $\pm$ 3.6	&	Kinematic distance	&	$v_{\scaleto{TP}{3pt}}$ - 0 	&	22	&	 \nodata     	&	 6.6       $\pm$ 1.5 	\\
71	&	       G$49.2-0.7$     	&	       5.4     $\pm$ 0.6	&	Kinematic distance	&	$v_{\scaleto{TP}{3pt}}$ 	&	22	&	     \nodata     	&	   6.3     $\pm$ 0.1 	\\
72	&	 G$51.04+0.07^\dagger $ 	&	      7.7 $\pm$ 2.3   	&	Kinematic distance	&	 $v_{\scaleto{TP}{3pt}}$	&	38	&	      7.2     $\pm$ 1.8 	&	 6.8 $\pm$ 0.5  	\\
73	&	 G$51.26+0.11^\dagger $ 	&	      \nodata  	&	\nodata	&	$v_{\scaleto{TP}{3pt}}$ - 37  	&	\nodata	&	 6.6 $\pm$ 1.7  	&	 6.6 $\pm$ 0.4 	\\
74	&	       G$53.4+0.0$     	&	7.5	&	Sagittarius–Carina arm	&	   \nodata 	&	39	&	       \nodata         	&	7.2	\\
75	&	       G$53.6-2.2$     	&	       2.3     $\pm$ 0.8       	&	Kinematic distance	&	27	&	40	&	   7.8     $\pm$ 0.6       	&	       7.3     $\pm$ 0.2 	\\
76	&	       G$54.1+0.3$     	&	       4.9     $\pm$ 0.8       	&	Kinematic distance	&	53.66	&	22	&	       \nodata         	&	       6.8     $\pm$ 0.1 	\\
77	&	       G$54.4-0.3$     	&	       6.6     $\pm$ 0.6       	&	Kinematic distance	&	36.66	&	31	&	       \nodata         	&	       7 $\pm$ 0.2       	\\
78	&	       G$55.0+0.3$     	&	14	&	Kinematic distance	&	   $-46.2$ 	&	41	&	       13.1 $\pm$ 0.7  	&	       10.7 $\pm$ 0.5	\\
79	&	       G$57.2+0.8$     	&	       6.6     $\pm$ 0.7       	&	Kinematic distance	&	       \nodata 	&	42	&	       \nodata         	&	       7.3     $\pm$ 0.2 \\
80	&	       G$59.5+0.1$     	&	2.3	&	Kinematic distance	&	28	&	43	&	2.7 $\pm$ 0.5 	&	7.3$ \pm$0.2	\\
81 &  G$63.7+1.1$ & 3.8 $\pm$ 1.5     &	Kinematic distance	&	18.7	&	44	&	       \nodata         	&	       7.5     $\pm$ 0.1 	\\
82	&	       G$64.5+0.9$     	&	11	&	Kinematic distance	&	       $-43$   	&	45	&	       10.6 $\pm$ 0.7  	&	       10.3 $\pm$ 0.5	\\
83	&	       G$65.1+0.6$     	&	       9.3     $\pm$ 0.3       	&	Kinematic distance	&	       $-23$   	&	45	&	       8.8 $\pm$ 0.6       	&	   9.2     $\pm$ 0.4 	\\
84	&	       G$65.3+5.7$     	&	       0.77 $\pm$ 0.2  	&	Proper motion + $v_{Exp}$	&	       \nodata 	&	46	&	       \nodata         	&	       8 $\pm$ 0.2   	\\
85	&	       G$65.7+1.2$     	&	       1 $\pm$ 0.4         	&	Kinematic distance	&	12	&	47	&	   1.3     $\pm$ 0.4       	&	       7.9     $\pm$ 0.2 	\\
86	&	 G$65.8-0.5^\ddagger$	&	 $2.4^{+0.3}_{-0.5}$ 	&	Optical extinction	&	 \nodata   	&	10	&	       \nodata         	&	 7.7   $\pm$ 0.3 	\\
87	&	       G$66.0+0.0$     	&	       2.3     $\pm$ 0.3       	&	Optical extinction	&	       \nodata 	&	10	&	       \nodata         	&	       7.7     $\pm$ 0.2 	\\
88	&	       G$67.6+0.9$     	&	       2 $\pm$ 0.2         	&	Optical extinction	&	   \nodata 	&	10	&	       \nodata         	&	       7.8     $\pm$ 0.2 	\\
89	&	       G$67.7+1.8$     	&	       $2^{+3.7}_{-0.5}$	&	Optical extinction	&	      \nodata 	&	10	&	       \nodata         	&	       $7.8^{+0.5}_{-0.2}$ 	\\
90	&	       G$69.0+2.7$     	&	1.5	&	Kinematic distance	&	12	&	48	&	       1.6     $\pm$ 0.7       	&	       7.9     $\pm$ 0.2 	\\
91	&	 G$73.9+0.9 $  	&	 4.4 $\pm$     0.1     	&	Kinematic distance	&	2.1	&	49	&	 4.2 $\pm$ 0.8         	&	 8.2 $\pm$ 0.2 	\\
92	&	 G$74.0-8.5 $  	&	 0.735 $\pm$ 0.025     	&	Parallax measurement	&	 \nodata       	&	50	&	       \nodata         	&	 8.2 $\pm$     0.2 	\\
93	&	 G$74.9+1.2 $  	&	 6.1 $\pm$     0.9         	&	Optical extinction	&	 \nodata   	&	51	&	       \nodata         	&	 9.0 $\pm$ 0.4 	\\
94	&	 G$75.2+0.1^\ddagger$	&	 $\geqslant$10 	&	Absorption column	&	 \nodata 	&	52	&	       \nodata         	&	 $\geqslant$11.2	\\
95	&	 G$76.9+1.0 $  	&	8	&	PWN 	&	 \nodata       	&	53	&	       \nodata         	&	10.2	\\
96	&	 G$78.2+2.1 $  	&	 2.1 $\pm$     0.4         	&	Kinematic distance	&	-8	&	54	&	   \nodata         	&	 8.6 $\pm$     0.2     	\\
97	&	 G$82.2+5.3 $  	&	 3.2 $\pm$     0.4         	&	Optical extinction	&	 \nodata   	&	10	&	       \nodata         	&	 8.5 $\pm$     0.2     	\\
98	&	 G$84.2-0.8 $  	&	 6     $\pm$   0.2         	&	Kinematic distance	&	-48	&	55	&	 7 $\pm$   0.7         	&	 10.3 $\pm$ 0.4	\\
99	&	 G$85.4+0.7 $  	&	 4.4 $\pm$     0.8         	&	Optical extinction	&	 \nodata   	&	10	&	       \nodata          	&	 9.1 $\pm$    0.4     	\\
100	&	 G$85.9-0.6 $  	&	 4.8 $\pm$     1.6         	&	Kinematic distance	&	32	&	56	&	   \nodata         	&	 9.3 $\pm$     0.7 	\\
101	&	 G$89.0+4.7 $  	&	 1.9$^{+0.3}_{-0.2}$	&	Optical extinction	&	  \nodata 	&	10	&	       \nodata         	&	 8.5 $\pm$ 0.2 	\\
102	&	 G$93.3+6.9 $  	&	 2.2 $\pm$     0.5         	&	HI column density	&	 \nodata   	&	57	&	       \nodata         	&	 8.7 $\pm$     0.2 	\\
103	&	 G$93.7-0.2 $  	&	 1.5 $\pm$     0.2         	&	Kinematic distance	&	-6	&	58	&	 1.3 $\pm$ 0.2         	&	 8.5 $\pm$ 0.2     	\\
104	&	 G$94.0+1.0 $  	&	3	&	Kinematic distance	&	-13	&	59	&	 2.2 $\pm$ 0.8         	&	 8.8 $\pm$ 0.3     	\\
105	&	 G$96.0+2.0 $  	&	4	&	Kinematic distance	&	-44	&	60	&	 4.9 $\pm$ 0.7         	&	 10.1 $\pm$ 0.4	\\
106	&	 G$106.3+2.7$  	&	0.8	&	Kinematic distance	&	-6.4	&	61	&	 0.7 $\pm$ 0.5     	&	 8.5 $\pm$     0.2 	\\
107	&	 G$107.0+9.0^\dagger$   	&	      1.75 $\pm$ 0.25 	&	Kinematic distance	&	 \nodata 	&	62	&	     \nodata     	&	 9 $\pm$   0.2  	\\
108	&	 G$107.5-1.5^\ddagger$	&	1.1	&	Kinematic distance	&	     $-9$    	&	63	&	0.9	&	8.6	\\
109	&	 G$108.2-0.6$  	&	 3.2 $\pm$     0.6         	&	Kinematic distance	&	-55	&	64	&	   \nodata          	&	 9.8 $\pm$    0.4     	\\
110	&	 G$109.1-1.0$  	&	 3.1 $\pm$     0.2         	&	Kinematic distance	&	$-50 \pm 6$	&	65	&	       \nodata         	&	 9.8 $\pm$     0.2     	\\
111	&	 G$111.7-2.1$  	&	 3.33 $\pm$ 0.1        	&	Proper motion	&	 -40$^{**}$	&	66	&	   3.3 $\pm$ 0.6 	&	 10 $\pm$ 0.4	\\
112	&	 G$113.0+0.2$  	&	3.1	&	Kinematic distance	&	-50	&	60	&	 3.2 $\pm$ 0.7         	&	 10 $\pm$  0.5 	\\
113	&	 G$114.3+0.3$  	&	0.7	&	Kinematic distance	&	-6.5	&	67	&	 0.5 $\pm$ 0.4         	&	 8.5 $\pm$ 0.2 	\\
114	&	 G$116.5+1.1$  	&	1.6	&	Kinematic distance	&	-17	&	67	&	 1.3 $\pm$ 0.6         	&	 9 $\pm$   0.4     	\\
115	&	 G$116.9+0.2$  	&	1.6	&	Kinematic distance	&	-27	&	67	&	 2.1 $\pm$     0.6         	&	 9.4 $\pm$ 0.4 	\\
116	&	 G$119.5+10.2$ 	&	 1.4 $\pm$     0.3         	&	Kinematic distance	&	-16	&	68	&	 1.2 $\pm$ 0.2         	&	 9.0 $\pm$ 0.2     	\\
117	&	 G$120.1+1.4$  	&	 4     $\pm$ 1         	&	Proper motion + $v_{Exp}$	&	 \nodata       	&	69	&	       \nodata         	&	 10.9 $\pm$ 0.8	\\
118	&	 G$126.2+1.6$  	&	 5.6 $\pm$     0.3         	&	Kinematic distance	&	 $-39$-$-45$	&	70	&	 3.2 $\pm$ 0.6    	&	 10.5 $\pm$ 0.5	\\
119	&	 G$127.1+0.5$  	&	1.15	&	Kinematic distance	&	-14	&	71	&	 0.96 $\pm$ 0.5    	&	 8.9 $\pm$     0.4     	\\
120	&	 G$130.7+3.1$  	&	 2     $\pm$ 0.3           	&	Perseus arm	&	 \nodata   	&	72	&	       \nodata         	&	 9.8 $\pm$     0.3     	\\
121	&	 G$132.7+1.3$  	&	 1.95 $\pm$ 0.04       	&	Kinematic distance	&	-45	&	73	&	   \nodata         	&	 9.8 $\pm$     0.2 	\\
122	&	 G$141.2+5.0^\ddagger$	&	 4 $\pm$ 0.5  	&	Kinematic distance	&	53	&	74	&	   \nodata         	&	 11.7  $\pm$ 0.5 	\\
123	&	 G$150.3+4.5$  	&	 0.7-4.5                       	&	Sedov estimates	&	 \nodata       	&	75	&	 \nodata           	&	 10.7 $\pm$ 1.8	\\
124	&	 G$152.4-2.1$  	&	 1.1 $\pm$     0.1         	&	Kinematic distance	&	12	&	76	&	 1.1 $\pm$ 0.4     	&	 9.3 $\pm$     0.4     	\\
125	&	 G$156.2+5.7$  	&	 $\gtrsim$1.7/ 1-3                    	&	Proper motion + $v_{Shock}$	&	 \nodata       	&	77	&	\nodata      	&	\nodata	\\
126	&	 G$159.6+7.3$  	&	 1-2.5                     	&	Diameter estimation	&	 \nodata   	&	78	&	 \nodata               	&	 10 $\pm$      0.7 	\\
127	&	 G$160.9+2.6$  	&	 0.8 $\pm$     0.4         	&	Kinematic distance	&	-6	&	79	&	 0.7 $\pm$ 0.4         	&	 9 $\pm$ 0.4   	\\
128	&	 G$166.0+4.3$  	&	 1     $\pm$ 0.4           	&	Kinematic distance	&	-6	&	80	&	       \nodata         	&	 9.3 $\pm$     0.4 	\\
129	&	 G$172.8+1.5^\ddagger$	&	1.8	&	HII distance	&	       $-20$   	&	81	&	       \nodata         	&	10.1	\\
130	&	 G$179.0+2.6$  	&	 $<5$                      	&	Estimation	&	 \nodata   	&	82	&	 \nodata                       	&	 $8.34<R<13.3$ 	\\
131	&	 G$180.0-1.7$	&	1.333$^{+0.103}_{-0.112}$	&	Kinematic distance	&	 \nodata 	&	83	&	 \nodata                     	&	 9.7$\pm$0.2	\\
132	&	 G$181.1+9.5$  	&	 1.5 $\pm$     1           	&	High V HI clouds	&	 \nodata   	&	84	&	       \nodata         	&	 9.8 $\pm$     1.0 	\\
133	&	 G$184.6-5.8$  	&	 3.37$^{+4.04}_{-0.97}$ 	&	Parallax measurement	&	 \nodata      	&	85	&	       \nodata         	&	       11.7$^{+4.04}_{-0.98}$ 	\\
134	&	 G$189.1+3.0$  	&	 1.729$^{+0.116}_{-0.094}$ 	&	MC association	&	 \nodata 	&	86	&	 \nodata         	&	 10.1 $\pm$ 0.2 	\\
135	&	 G$189.6+3.3^\ddagger$	&	 1.5 $ \pm$     0.2     	&	Object associtaions	&	 \nodata 	&	87	&	     \nodata         	&	 9.8   $\pm$ 0.3                 	\\
136	&	 G$190.9-2.2 $ 	&	 1.036$^{+0.079}_{-0.087}$ 	&	MC association	&	 \nodata 	&	86	&	 \nodata         	&	       9.4     $\pm$ 0.2         	\\
137	&	 G$205.5+0.5 $ 	&	 1.15 $\pm$ 0.3            	&	MC association	&	 \nodata 	&	86	&	 \nodata         	&	 9.4 $\pm$     0.3               	\\
138	&	 G$206.9+2.3 $ 	&	2.2	&	Kinematic distance	&	$22 \pm 5$	&	88	&	     2.2     $\pm$ 0.7       	&	 10.3 $\pm$ 0.7        	\\
139	&	 G$213.0-0.6 $ 	&	 1.146$^{+0.079}_{-0.080}$     	&	MC association	&	 \nodata 	&	86	&	     \nodata         	&	       9.4     $\pm$ 0.2 	\\
140	&	 G$260.4-3.4 $ 	&	 1.3 $\pm$     0.3                 	&	Kinematic distance	&	10	&	89	&	     1.4     $\pm$ 0.8       	&	 8.7 $\pm$     0.3               	\\
141	&	 G$263.9-3.3 $ 	&	 0.25 $\pm$ 0.03               	&	Parallax measurement	&	 \nodata 	&	90	&	     \nodata         	&	 8.4 $\pm$     0.2                       	\\
142	&	 G$266.2-1.2 $ 	&	 0.7 $\pm$ 0.2               	&	MC association	&	 \nodata 	&	91	&	     \nodata         	&	 8.4 $\pm$     0.2                       	\\
143	&	 G$267.0-1.0^\ddagger$	&	 $\leqslant$0.9 	&	MC association	&	 \nodata 	&	92	&	     \nodata         	&	 8.39  $\pm$ 0.05	\\
144	&	 G$272.2-3.2 $ 	&	$2$- $10$	&	Stats + optical color excess	&	 \nodata 	&	93	&	     \nodata         	&	 10.1 $\pm$ 2.3                  	\\
145	&	 G$276.5+19.0^\ddagger$	&	 0.2 $ \pm$ 0.14 	&	$^{26}$Al  $\gamma$-ray emission	&	 \nodata 	&	94	&	   \nodata         	&	 8.3   $\pm$ 0.2 	\\
146	&	 G$279.0+1.1 $ 	&	 2.7 $\pm$     0.3                 	&	Optical extinction	&	 \nodata 	&	95	&	 \nodata         	&	 8.4 $\pm$     0.2                       	\\
147	&	 G$284.3-1.8 $ 	&	 5.5 $\pm$     0.7                 	&	Optical extinction	&	 \nodata 	&	95	&	 \nodata         	&	 8.8 $\pm$     0.3               	\\
148	&	 G$290.1-0.8 $ 	&	 7     $\pm$ 1                 	&	Kinematic distance	&	7	&	96	&	 6.3 $\pm$ 0.8   	&	 8.9 $\pm$     0.5                       	\\
149	&	 G$291.0-0.1 $ 	&	3.5	&	HII distance	&	 \nodata 	&	97	&	 \nodata         	&	7.8	\\
150	&	 G$292.0+1.8 $ 	&	 6.2 $\pm$     0.9                 	&	Kinematic distance	&	0	&	98	&	 6.2     $\pm$ 0.8       	&	 8.6 $\pm$     0.3                       	\\
151	&	 G$292.2-0.5 $ 	&	 8.4 $\pm$     0.4                 	&	Kinematic distance	&	22	&	99	&	     8.1     $\pm$ 0.6       	&	 9.2 $\pm$     0.3                       	\\
152	&	 G$296.1-0.5 $ 	&	 4.3 $\pm$     0.8                 	&	Optical extinction	&	 \nodata 	&	95	&	 \nodata         	&	 7.5 $\pm$     0.2                       	\\
153	&	 G$296.5+10.0$ 	&	 2.1$^{+1.8}_{-0.8}$   	&	Kinematic distance	&	-16	&	100	&	     \nodata         	&	 7.6$^{+0.4}_{-0.2}$	\\
154	&	 G$296.7-0.9 $ 	&	  9.8$^{+1.1}_{-0.7}$                	&	Sedov estimates	&	 \nodata 	&	101	&	 \nodata         	&	 9.7 $\pm$     0.6                       	\\
155	&	 G$296.8-0.3 $ 	&	 9.6 $\pm$     0.6                 	&	Kinematic distance	&	23	&	102	&	     9.3     $\pm$ 0.6       	&	 9.3 $\pm$     0.4               	\\
156	&	 G$298.6+0.0 $ 	&	10	&	Absorption column	&	 \nodata 	&	103	&	 \nodata         	&	9.5	\\
157	&	 G$299.2-2.9 $ 	&	 2.8 $\pm$     0.8                             	&	Optical extinction	&	 \nodata 	&	95	&	     \nodata         	&	 7.4 $\pm$     0.2               	\\
158	&	 G$304.6+0.1 $ 	&	9.7	&	Kinematic distance	&	-20	&	104	&	     7.9     $\pm$ 0.6       	&	 7.5 $\pm$     0.3                       	\\
159	&	 G$306.3-0.9 $ 	&	20	&	Absorption column	&	 \nodata 	&	105	&	     \nodata         	&	16.5	\\
160	&	 G$308.4-1.4 $ 	&	 5     $\pm$ 0.7                               	&	Optical extinction	&	 \nodata 	&	95	&	     \nodata         	&	 6.5 $\pm$     0.1               	\\
161	&	 G$308.8-0.1 $ 	&	5.4	&	Estimation	&	 \nodata 	&	106	&	             \nodata     	&	6.5	\\
162	&	 G$309.2-0.6 $ 	&	 2.8 $\pm$     0.8                             	&	Optical extinction	&	 \nodata 	&	95	&	     \nodata         	&	 6.9 $\pm$     0.3               	\\
163	&	 G$309.8-2.6^\ddagger$	&	 2.3 $ \pm$     0.2     	&	Optical extinction	&	 \nodata 	&	95	&	     \nodata         	&	 7.1   $\pm$ 0.2 	\\
164	&	 G$310.6-1.6 $ 	&	7	&	Crux–Scutum arm	&	 \nodata 	&	107	&	     \nodata         	&	6.5	\\
165	&	 G$311.5-0.3 $ 	&	 $>6.6$                                        	&	Kinematic distance	&	-10	&	104	&	     10.3 $\pm$ 0.5  	&	 7.9 $\pm$     0.3               	\\
166	&	 G$312.4-0.4 $ 	&	 $>6$                                          	&	Kinematic distance	&	-50	&	108	&	 3.5 $\pm$ 0.5  	&	 6.5 $\pm$     0.2               	\\
167	&	 G$313.3+0.1^\ddagger$	&	5.6	&	PWN	&	 \nodata     	&	53	&	 \nodata         	&	6.1	\\
168	&	 G$315.1+2.7 $ 	&	1.7	&	Estimation	&	 \nodata 	&	109	&	     \nodata         	&	7.2	\\
169	&	 G$315.4-2.3 $ 	&	 2.8 $\pm$     0.4                             	&	Kinematic distance	&	-33.2	&	110	&	     2.2     $\pm$ 0.4       	&	 6.9 $\pm$     0.3               	\\
170	&	 G$316.3+0.0 $ 	&	 $>7.2$                                                	&	Kinematic distance	&	-40	&	104	&	     9.4     $\pm$ 0.4       	&	 6.7 $\pm$     0.2       	\\
171	&	 G$318.2+0.1 $ 	&	 3.5 $\pm$     0.2                             	&	Kinematic distance	&	-42	&	111	&	     2.7     $\pm$ 0.4 	&	 6.6 $\pm$ 0.3                     	\\
172	&	 G$320.4-1.2 $ 	&	 5.2 $\pm$     1.4                             	&	Kinematic distance	&	 $-55$-$-70$ 	&	112	&	 \nodata     	&	    5.5 $\pm$      0.3       	\\
173	&	 G$322.1+0.0 $ 	&	 9.4$^{+0.8}_{-1}$             	&	Kinematic distance	&	 \nodata 	&	113	&	     \nodata          	&	      5.8$^{+0.4}_{-0.5}$       	\\
174	&	 G$323.7-1.0 $ 	&	3.5	&	Scutum-Crux arm	&	 \nodata 	&	114	&	     \nodata         	&	5.9	\\
175	&	 G$326.3-1.8 $ 	&	 4.1 $\pm$     0.7                             	&	Kinematic distance	&	-58	&	110	&	     3.5     $\pm$ 0.6       	&	 5.8 $\pm$     0.4                       	\\
176	&	 G$327.1-1.1 $ 	&	9	&	Absorption column	&	 \nodata 	&	115	&	     \nodata         	&	5	\\
177	&	 G$327.2-0.1 $ 	&	 4.5 $\pm$     0.5                             	&	CO observations	&	 \nodata 	&	116	&	     \nodata         	&	 5.2 $\pm$     0.3                       	\\
178	&	 G$327.4+0.4 $ 	&	 4.3 $\pm$     0.5                             	&	Kinematic distance	&	-75	&	117	&	     4.4     $\pm$ 0.5       	&	 5.2 $\pm$     0.3     	\\
179	&	 G$327.6+14.6 $ 	&	 2.18 $\pm$ 0.08	&	Proper motion + $v_{Shock}$	&	\nodata      	&	118	&	       \nodata         	&	 6.6 $\pm$ 0.2  	\\
180	&	 G$328.4+0.2  $ 	&	 $>17.4 \pm$ 0.9	&	Kinematic distance	&	28	&	119	&	   $ >16.7 \pm$ 0.7        	&	 $>10.5 \pm 0.6$ 	\\
181	&	 G$330.0+15.0 $ 	&	1.2	&	Estimation	&	     \nodata 	&	120	&	       \nodata         	&	7.3	\\
182	&	 G$330.2+1.0  $ 	&	 $>4.9 \pm$ 0.3 	&	Kinematic distance	&	     \nodata 	&	117	&	       \nodata         	&	 $>4.1 \pm$ 0.2 	\\
183	&	 G$332.4-0.4  $ 	&	 3.0 $\pm$ 0.3  	&	Optical extinction	&	     \nodata 	&	95	&	       \nodata         	&	 5.8 $\pm$     0.3  	\\
184	&	 G$332.4+0.1  $ 	&	 9.2 $\pm$ 1.7  	&	Norma arm	&	     \nodata 	&	121	&	       \nodata         	&	 4.3 $\pm$     0.7  	\\
185	&	 G$332.5-5.6  $ 	&	 3 $\pm$ 0.8    	&	Optical extinction	&	     \nodata 	&	122	&	       \nodata         	&	 5.8 $\pm$     0.6  	\\
186	&	 G$335.2+0.1  $ 	&	 1.8 $\pm$ 0.4  	&	Kinematic distance	&	-22.5	&	123	&	       1.7     $\pm$ 0.4       	&	 6.8 $\pm$     0.4      	\\
187	&	 G$337.0-0.1  $ 	&	 11 $\pm$ 0.3   	&	Kinematic distance	&	-72	&	124	&	   10.8 $\pm$ 0.3  	&	 4.5 $\pm$     0.2  	\\
188	&	 G$337.2-0.7  $ 	&	 5.5 $\pm$ 4    	&	Kinematic distance	&	-100	&	125	&	       9.4     $\pm$ 0.3       	&	 3.7 $\pm$     0.1  	\\
189	&	 G$337.2+0.1  $ 	&	14	&	Kinematic distance	&	     \nodata 	&	126	&	       \nodata         	&	7.1	\\
190	&	 G$337.8-0.1  $ 	&	12.3	&	Kinematic distance	&	-45	&	127	&	   12.2 $\pm$ 0.3  	&	 5.5 $\pm$     0.3      	\\
191	&	 G$338.3+0.0  $ 	&	 8.5-13                 	&	Kinematic distance	&	-31	&	128	&	   13 $\pm$ 0.4    	&	 6.1 $\pm$     0.4      	\\
192	&	 G$338.5+0.1  $ 	&	11	&	Norma arm	&	     \nodata 	&	129	&	       \nodata         	&	4.5	\\
193	&	 G$340.6+0.3  $ 	&	15	&	Scutum-Crux arm	&	     \nodata 	&	129	&	       \nodata         	&	7.7	\\
194	&	 G$341.2+0.9  $ 	&	6.9	&	Pulsar association	&	     \nodata 	&	130	&	       \nodata         	&	2.9	\\
195	&	 G$341.9-0.3  $ 	&	 \nodata                	&	\nodata	&	0	&	127	&	   15.8 $\pm$ 0.6  	&	 8.3 $\pm$ 0.6  	\\
196	&	 G$342.0-0.2  $ 	&	 \nodata                	&	\nodata	&	0	&	127	&	   15.8 $\pm$ 0.6  	&	 8.3 $\pm$ 0.6  	\\
197	&	 G$343.0-6.0  $ 	&	$1.2 \pm 0.2$	&	Civ luminosity estimation	&	     \nodata 	&	131, 132	&	       \nodata         	&	7.4	$\pm $ 0.2 \\
198	&	 G$343.1-0.7  $ 	&	 \nodata                	&	\nodata	&	-70	&	127	&	   4.9     $\pm$ 0.2       	&	 3.9 $\pm$ 0.2  	\\
199	&	 G$344.7-0.1  $ 	&	 6.3  $\pm$ 0.1 	&	Kinematic distance	&	-115	&	133	&	       \nodata         	&	 2.8 $\pm$     0.1  	\\
200	&	 G$346.6-0.2  $ 	&	 5.5/11                 	&	Kinematic distance	&	-76	&	127	&	   10.4 $\pm$ 0.2  	&	 3.0 $\pm$ 0.2  	\\
201	&	 G$347.3-0.5  $ 	&	1	&	Kinematic distance	&	-6	&	134	&	   0.9     $\pm$ 0.6       	&	 7.5 $\pm$     0.6  	\\
202	&	 G$348.5+0.0  $ 	&	 $\leqslant6.3$ 	&	Kinematic distance	&	     -100$^{**}$     	&	135	&	 7.2 $\pm$ 0.2 	&	 1.9 $\pm$     0.1  	\\
203	&	 G$348.5+0.1  $ 	&	 $6.3$ - $9.5$	&	Kinematic distance	&	     $-107$-$-145$  	&	135	&	        \nodata         	&	 1.7 $\pm$     0.3      	\\
204	&	 G$348.7+0.3  $ 	&	 13.2 $\pm$ 0.2 	&	Kinematic distance	&	26	&	135	&	   \nodata         	&	 5.3 $\pm$     0.2  	\\
205	&	 G$349.7+0.2  $ 	&	 11.5 $\pm$ 0.7 	&	3 kpc arm	&	     \nodata 	&	136	&	       \nodata         	&	 3.6 $\pm$     0.7      	\\
206	&	 G$350.0-2.0  $ 	&	3	&	Optical extinction	&	     \nodata 	&	137	&	       \nodata         	&	5.4	\\
207	&	 G$350.1-0.3  $ 	&	 9    $\pm$ 3   	&	Absorption column	&	     \nodata 	&	138	&	       \nodata         	&	 1.6 $\pm$     1.4  	\\
208	&	 G$351.7+0.8  $ 	&	 13.2 $\pm$ 0.5 	&	Kinematic distance	&	-14	&	139	&	   14 $\pm$ 0.6    	&	 5.9 $\pm$     0.6  	\\
209	&	 G$352.7-0.1  $ 	&	 7.5  $\pm$ 0.5 	&	Kinematic distance	&	     \nodata 	&	140	&	       \nodata         	&	 1.3 $\pm$     0.3  	\\
210	&	 G$353.6-0.7  $ 	&	 3.2  $\pm$ 0.8 	&	Kinematic distance	&	-30	&	141	&	   4.8     $\pm$ 0.8       	&	 3.6 $\pm$     0.8      	\\
211	&	 G$354.4+0.0^\ddagger$	&	 5 - 8                  	&	Kinematic distance	&	 \nodata 	&	142	&	 \nodata      	&	 2 $\pm$ 0.2    	\\
212	&	 G$354.8-0.8  $ 	&	 \nodata                	&	\nodata	&	-70	&	127	&	   8.3     $\pm$ 0.2       	&	1	\\
213	&	 G$355.6+0.0  $ 	&	13	&	Absorption column	&	     \nodata 	&	143	&	       \nodata         	&	4.7	\\
214	&	 G$357.7-0.1  $ 	&	11.8	&	Kinematic distance	&	-12.4	&	144	&	       11.5 $\pm$ 0.7  	&	 3.2 $\pm$ 0.7  	\\
215	&	 G$359.1-0.5  $ 	&	5	&	Kinematic distance	&	     \nodata 	&	145	&	       \nodata         	&	3.3	\\
\enddata
\tablecomments{a - $\dagger$ indicates new SNRs that are not in the catalogue presented by \cite{2019GreenCat} and $\ddagger$ indicates uncertain SNRs (I.e. more observational evidence is needed). \\
b - The distances for most SNRs and the errors are recalculated using the rotaion curve parameters presented by \cite{2014ReidRC}. \\ 
$^{*}$- The $V_{LSR} $ is from \cite{2009Hewitt} \\
$^{**}$- The $V_{LSR} $ is from \cite{1998Koralesky}\\ 
References : (1)  \cite{1993Reid}, (2)  \cite{1996KassimFrail},  (3)  \cite{2005Aharonian}, (4)   \cite{2014Philstrom}, (5)   \cite{2008Reynolds}, (6)  \cite{2016Sankrit}, (7)   \cite{2009Hewitt}, (8) \cite{1994FrailKassimWeiler}, (9)  \cite{2009Liszt}, (10) \cite{2018Shan},  (11)  \cite{2002Velazquez}, (12)  \cite{2008RobertsBrogen}, (13)  \cite{1986Milne}, (14)  \cite{2016Kilpatrick}, (15)  \cite{2014Yamauchi}, (16)  \cite{2009Gotthelf}, (17) \cite{1995Seward},   (18)  \cite{2017Su},  \cite{2019Tian},  (20)  \cite{2011Beaumont}, (21)  \cite{2002Boumis}, (22) \cite{2018RanasingheApJ}, (23)  \cite{2021Ranasinghe}, (24) \cite{2020RanasJHEGC}, (25) \cite{2020Gao} , (26)  \cite{2018RanasMNRAS}, (27)  \cite{2018RanasOPhyJ}, (28) \cite{2011Gao}, (29) \cite{2003Bamba} , (30) \cite{2019Petriella}, (31)  \cite{2017Ranasinghe}, (32)  \cite{2014Huang},  (33) \cite{2018Su}, (34)  \cite{2006Yang}, (35)  \cite{2000Lorimer}  (36) \cite{2020Zhou}, (37)  \cite{2017SuYangJi}, (38)  \cite{2018Supan}, (39) \cite{2018Driessen}, (40)  \cite{1998Giacani}, (41) \cite{1998Matthews},  (42)  \cite{2020Zhou1}, (43)  \cite{2012Xuwang}, (44) \cite{1997Wallace}, (45)  \cite{2006TianLeahy}, (46)  \cite{2004Boumis}, (47)  \cite{2008Kothes}, (48)  \cite{2012LeahyRanasinghe}, (49) \cite{2016Zdziarski}, (50)  \cite{2018Fesen}, (51)  \cite{2003Kothes}, (52) \cite{2004Hessels}, (53)   \cite{2010Kargaltsev}, (54)  \cite{2013LeahyGreenRana} , (55)  \cite{2012LeahyGreen}, (56)  \cite{2008Jackson}, (57)   \cite{2003FosterRout} , (58) \cite{2002Uyaniker}, (59) \cite{2013Jeong},  (60) \cite{2005KothesUyan}, (61)  \cite{2001KothesUyaniker}, (62) \cite{2020Fesen}, (63)  \cite{2003KothesR107}, (64)  \cite{2007TianLeahyFos}, (65)  \cite{2018SanchezCruces}, (66)  \cite{2014Alarie}, (67)  \cite{2004YarUyaniker}, (68) \cite{1993Pineault}, (69)  \cite{2010Hayato}, (70)  \cite{2006TianLeahy2}, (71)  \cite{2006LeahyTian2}, (72)  \cite{2013Kothes}, (73)  \cite{2016Zhou}, (74)  \cite{2014Kothessun}, (75)  \cite{2020Devin}, (76)  \cite{2013Foster}, (77)   \cite{2016Katsuda}, (78)   \cite{2010FesenMil}, (79)  \cite{2007LeahyTian2}, (80)   \cite{2019Arias}, (81)  \cite{2012Kang}, (82)   \cite{2018How}, (83)  \cite{2015Dincel}, (84) \cite{2017KothesReich}, (85)  \cite{2019FraB}, (86) \cite{2019YuChen}, (87)  \cite{1994Asaoka}, (88) \cite{2014AmbrocioCruz}, (89) \cite{2017Reynoso}, (90) \cite{1999Cha}, (91) \cite{2015Allen}, (92) \cite{2013Acero},  (93)  \cite{2012Sezer}, (94) \cite{2002McCullough}, (95) \cite{2019Shan}, (96)  \cite{2006Reynoso}, (97) \cite{1986Roger}, (98) \cite{2003Gaensler},  (99) \cite{2004Caswell}, (100)  \cite{2000Giacani}, (101)  \cite{2013Prinz}, (102)  \cite{1998Gaensler}, (103)  \cite{2016Bamba}, (104)  \cite{1975Caswell}, (105)  \cite{2019Sawada}, (106) \cite{1986Wilson}, (107)   \cite{2010Renaud}, (108)   \cite{2003Doherty}, (109) \cite{2007Stupar}, (110)  \cite{1996Rosado}, (111)  \cite{2010Hofverberg}, (112) \cite{1999Gaensler2}, (113)  \cite{2015Heinz}, (114) \cite{2018Maxted}, (115)  \cite{1999Sun}, (116)  \cite{2010Tiengo}, (117) \cite{2001McClureGriffiths}, (118)   \cite{2003Winkler}, (119)  \cite{2000Gaensler}, (120) \cite{1991Leahy}, (121)   \cite{2004Vink}, (122) \cite{2015Zhu}, (123)  \cite{2011Eger}, (124) \cite{1997Sarma}, (125) \cite{2006Rakowski}, (126) \cite{2005Combi}, (127) \cite{1998Koralesky}, (128) \cite{2016Supan}, (129) \cite{2007Kothes}, (130) \cite{1994Frail},  (131)  \cite{2003Welsh}, (132)  \cite{2010Kim}, (133) \cite{2011Giacani}, (134)  \cite{2003Fukui}, (135)  \cite{2012Tian}, (136) \cite{2014Tian}, (137) \cite{2016Karpova},  (138) \cite{2014Yasumi}, (139)  \cite{2007TianHav}, (140) \cite{2009Giacani}, (141) \cite{2008TianLeahyHav}, (142)  \cite{2013RoyPal}, (143)  \cite{2013Minami}, (144)  \cite{1996FrailGoss},  (145) \cite{2011Frail}. 
 }
\end{deluxetable*}

\begin{deluxetable*}{lllcc}
\tablenum{3}
\label{tab:3}
\tablecaption{SNR Surface Density and Correction Factor Model Parameters from Method 2}
\tablewidth{0pt}
\tablehead{
\colhead{\#} & \colhead{Models} &  \colhead{Parameters}          & \colhead{Y$^2$}  & \colhead{\# of SNRs$^a$} \\
\colhead{}   & \colhead{}       &  \colhead{}                    & \colhead{}       & \colhead{}         
}
\startdata
& & \\
1  & Exp - Exp  &  $A = 71.0 \pm 20.0 $ kpc$^{-2}$, $\quad$ $H_{\textrm{R}}  = 3.41^{+0.42}_{-0.45}$ kpc,    & $166.76$ & $4950^{+2840}_{-2210}$ \\
   &            & $H_{\textrm{S}}  = 1.95^{+0.30}_{-0.33}$ kpc,$\quad$ $  B = 0.60^{+0.24}_{-0.19}$, $\quad$ $l_0 =  131\degr^{+33\degr}_{-27\degr}$ &\\ 
\hline
2  & Exp - PL$_{\textrm{C}}$ &  $A = 42.2 \pm 12.9 $ kpc$^{-2}$, $\quad$ $H_{\textrm{R}}  = 4.99^{+1.03}_{-0.97}$ kpc,    & $165.06$ & $5590^{+3990}_{-2730}$ \\ 
   &            & $R_{\textrm{S}}  = 3.60^{+0.74}_{-0.78}$ kpc,$\quad$ $\alpha_2 = 3.18^{+0.57}_{-0.38} $, $\quad $ $  B = 0.49^{+0.33}_{-0.23}$, $\quad$ $l_0 =  116\degr^{+49\degr}_{-42\degr}$ &\\ 
\hline    
3  & PL${_\textrm{C}}^b$ - Exp &  $A = 45.0 \pm 13.1 $ kpc$^{-2}$, $\quad$ $R_{\textrm{c}}  = 4.29^{+1.02}_{-1.00}$ kpc, $\quad$ $\alpha_1 = 2 $,      & $168.31$ & $4120^{+2700}_{-2020}$ \\
   &            & $H_{\textrm{S}}  = 2.06^{+0.34}_{-0.37} $ kpc, $\quad $ $B = 0.52^{+0.38}_{-0.26} $, $\quad$ $l_0 =  98\degr^{+43\degr}_{-41\degr}$ &\\ 
\hline   			
4  & PL${_{\textrm{C}}}^b$ - PL${_\textrm{C}}$ & $A = 31.2 \pm 9.6$ kpc$^{-2}$, $\quad$ $R_{\textrm{c}}  = 7.83^{+2.75}_{-2.35}$ kpc, $\quad$ $\alpha_1 = 2 $,   & $166.31$ & $5530^{+4060}_{-2910}$ \\
   &            & $R_{\textrm{S}}  = 3.47^{+0.74}_{-0.77} $ kpc, $\quad$  $ \alpha_2 =  3.04^{+0.56}_{-0.37}$, $\quad$ $B = 0.46^{+0.44}_{-0.27} $, $\quad$ $l_0 =  90\degr^{+54\degr}_{-53\degr}$ &\\
\hline   
5  & PL${_{\textrm{C}}}^c$ - Exp&  $A = 52.1 \pm 16.2 $ kpc$^{-2}$, $\quad$ $R_{\textrm{c}}  = 2.26^{+0.64}_{-0.62}$ kpc, $\quad$ $\alpha_1 = 1.5 $,      & $168.37$ & $4150^{+2940}_{-2150}$ \\
   &            & $H_{\textrm{S}}  = 2.06^{+0.36}_{-0.40} $ kpc, $\quad $ $B = 0.52^{+0.44}_{-0.28} $, $\quad$ $l_0 =  93\degr^{+47\degr}_{-45\degr}$ &\\
\hline    
6  & PL${_{\textrm{C}}}^c$ - PL${_\textrm{C}}$ &  $A = 30.0 \pm 9.3$ kpc$^{-2}$, $\quad$ $R_{\textrm{c}}  = 5.21^{+2.06}_{-1.72}$ kpc, $\quad$ $\alpha_1 = 1.5 $,   & $166.57$ & $5510^{+4040}_{-2900}$ \\
   &            & $R_{\textrm{S}}  = 3.52^{+0.75}_{-0.79} $ kpc, $\quad$  $ \alpha_2 =  3.05^{+0.56}_{-0.37}$, $\quad$ $B = 0.46^{+0.46}_{-0.28} $, $\quad$ $l_0 =  85\degr \pm 54\degr$ &\\
\hline    
7 & GD - Exp & $A = 14.0 \pm 4.2 $ kpc$^{-2}$, $\quad$ $\mu  = 3.66^{+0.78}_{-0.95}$ kpc, $\quad$ $\sigma = 3.37 \pm 0.55 $ kpc,  & $163.95$ & $2900^{+2370}_{-1610}$ \\
   &            & $H_{\textrm{S}}  = 2.54^{+0.55}_{-0.56} $ kpc, $\quad $ $B = 0.62 \pm 0.26 $, $\quad$ $l_0 =  145\degr^{+32\degr}_{-28\degr}$ & \\
\hline
8 & GD - PL${_\textrm{C}}$ & $A = 19.3 \pm 6.2$ kpc$^{-2}$, $\quad$ $\mu  = 4.10^{+0.89}_{-1.08}$ kpc, $\quad$ $\sigma = 3.32^{+0.69}_{-0.66}$ kpc,   & $161.07$ & $4310^{+4000}_{-2550}$ \\
   &            & $R_{\textrm{S}}  = 2.46 \pm 0.66 $ kpc, $\quad$  $ \alpha_2 =  2.22^{+0.46}_{-0.30}$, $\quad$ $B = 0.57 \pm0.30 $, $\quad$ $l_0 =  144\degr^{38}_{33}\degr$ &\\
\hline   
9 & GD$_{\mu = 0}$ - Exp & $A = 32.2 \pm 8.9 $ kpc$^{-2}$, $\quad$ $\sigma = 4.56^{+0.40}_{-0.43} $ kpc,  & $164.58$ & $4200^{+2110}_{-1700}$ \\
   &            & $H_{\textrm{S}}  = 2.19^{+0.37}_{-0.40} $ kpc, $\quad $ $B = 0.66 \pm 0.22 $, $\quad$ $l_0 =  145\degr^{+27\degr}_{-22\degr}$ & \\
\hline   
10 & GD$_{\mu = 0}$ - PL${_\textrm{C}}$ & $A = 32.4 \pm 9.7$ kpc$^{-2}$, $\quad$ $\sigma = 5.10^{+0.61}_{-0.63}$ kpc,   & $162.44$ & $5270^{+3230}_{-2410}$ \\
   &            & $R_{\textrm{S}}  = 3.20^{+0.69}_{-0.72} $ kpc, $\quad$  $ \alpha_2 =  2.82^{+0.51}_{-0.34}$, $\quad$ $B = 0.57 \pm 0.29 $, $\quad$ $l_0 =  139\degr^{37}_{31}\degr$ &\\
\hline   
11  & MGF - Exp &  $A = 5.7 \pm 1.7 $ kpc$^{-2}$, $\quad$ $\alpha_1  = 3.71^{+1.26}_{-0.71}$ kpc, $\quad$ $\beta = 7.50^{+0.98}_{-1.43} $,  & $165.04$ & $3320^{+5210}_{-1640}$ \\
   &            & $H_{\textrm{S}}  = 2.38^{+0.51}_{-0.52} $ kpc, $\quad $ $B = 0.68^{+0.20}_{-0.16} $, $\quad$ $l_0 =  148\degr^{+28\degr}_{-25\degr}$ & \\
\hline    
12 & MGF - PL${_\textrm{C}}$&   $A = 8.9 \pm 2.9$ kpc$^{-2}$, $\quad$ $\alpha_1  = 3.70^{+1.33}_{-0.71}$ kpc, $\quad$ $\beta = 7.10^{+0.99}_{-1.51} $,   & $162.15$ & $4790^{+6910}_{-2010}$ \\
   &            & $R_{\textrm{S}}  = 2.32^{+0.61}_{-0.62} $ kpc, $\quad$  $ \alpha_2 =  2.23^{+0.46}_{-0.30}$, $\quad$ $B = 0.60^{+0.24}_{-0.18} $, $\quad$ $l_0 =  146\degr^{36\degr}_{31\degr}$ &\\
\hline   
13 & CB98e15 - Exp &   $A = 17.7 \pm 5.8 $ kpc$^{-2}$, $\quad$ $R_{\textrm{a}}  = 15.70^{+2.62}_{-1.78}$ kpc, $\quad$ $\theta_0 = 0.69^{+0.25}_{-0.29} $, $\quad$ $\beta = 0.11^{+0.05}_{-0.04}$, & $163.69$ & $2570^{+3960}_{-1630}$ $^*$\\
   &            &   $H_{\textrm{S}}  = 2.56^{+0.59}_{-0.61} $ kpc, $\quad $ $B = 0.57^{+0.28}_{-0.21} $, $\quad$ $l_0 =  135\degr^{+40\degr}_{-35\degr}$ & \\
\hline 
14 & CB98e15 - PL${_\textrm{C}}$&    $A = 28.8 \pm 9.9 $ kpc$^{-2}$, $\quad$ $R_{\textrm{a}}  = 14.02^{+3.33}_{-1.87}$ kpc, $\quad$ $\theta_0 = 0.35^{+0.33}_{-0.42} $, $\quad$ $\beta = 0.13^{+0.06}_{-0.04}$,  & $161.16$ & $3810^{+6310}_{-2470}$ $^*$ \\
   &            & $R_{\textrm{S}}  = 2.77 \pm 0.78 $ kpc, $\quad$ $\alpha_2 = 2.36^{+0.55}_{-0.34}$, $\quad $ $B = 0.52^{+0.30}_{-0.22} $, $\quad$ $l_0 =  136\degr^{+47\degr}_{-41\degr}$ & \\
\hline
15 & S{\'e}rsic - Exp&  $A = 8.3 \pm 2.5 $ kpc$^{-2}$, $\quad$ $R_{\textrm{e}}  = 6.24^{+0.62}_{-0.65}$ kpc, $\quad$ $n = 0.25^{+0.52}_{-0.16} $, $\quad$ $b_n = 0.23\quad(^{1.22}_{0.02})^d$,  & $163.42$ & $2380^{+2780}_{-1210}$ \\
   &            & $H_{\textrm{S}}  = 2.72 \pm 0.61 $ kpc, $\quad $ $B = 0.56 \pm 0.30$, $\quad$ $l_0 =  138\degr^{+37\degr}_{-33\degr}$ &\\
\hline 
16 & S{\'e}rsic - PL${_\textrm{C}}$& $A = 13.0 \pm 4.1 $ kpc$^{-2}$, $\quad$ $R_{\textrm{e}}  = 6.3 \pm 0.7$ kpc, $\quad$ $n = 0.25^{+0.50}_{-0.17} $, $\quad$ $b_n = 0.23\quad(^{1.20}_{0.01})^d$,  & $160.94$ & $3790^{+4680}_{-2050}$ \\
   &            & $R_{\textrm{S}}  = 2.80 \pm 0.75 $ kpc, $\quad $ $\alpha_2 = 2.32^{+0.50}_{-0.32}$,$\quad $  $B = 0.54 \pm 0.32 $, $\quad$ $l_0 =  140\degr^{+42\degr}_{-37\degr}$ &\\
\hline
\enddata
\tablecomments{a - The total number of SNRs estimated to galactocentric distance, $R = 16.68 $ kpc.
b,c - SNR surface density, $\rho_{\scaleto{SNR}{3pt}}(R)$ follows a power-law distribution where $\alpha_1 $ is set to 2 and 1.5, respectively. \\
$^{*}$ - The total number of SNRs are calculated up to a $R < R_a(1 - \theta_0/\pi)$\\
d - $b_n$ values calculated by using the lower and upper limits of $n$}
\end{deluxetable*}

\begin{figure*}
\centering
\includegraphics[width=16cm]{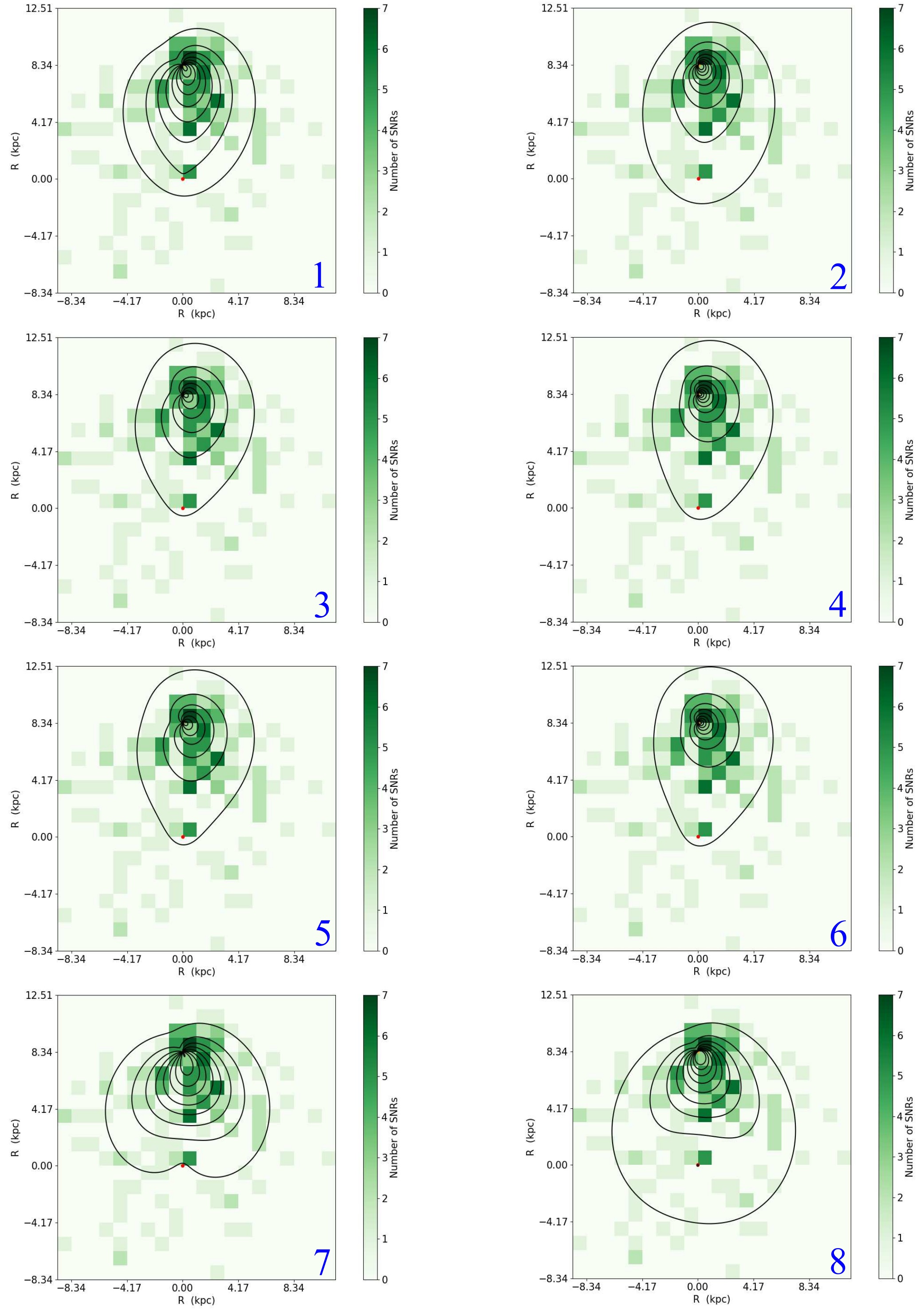}
\end{figure*}

\begin{figure*}
\centering
\includegraphics[width=16cm]{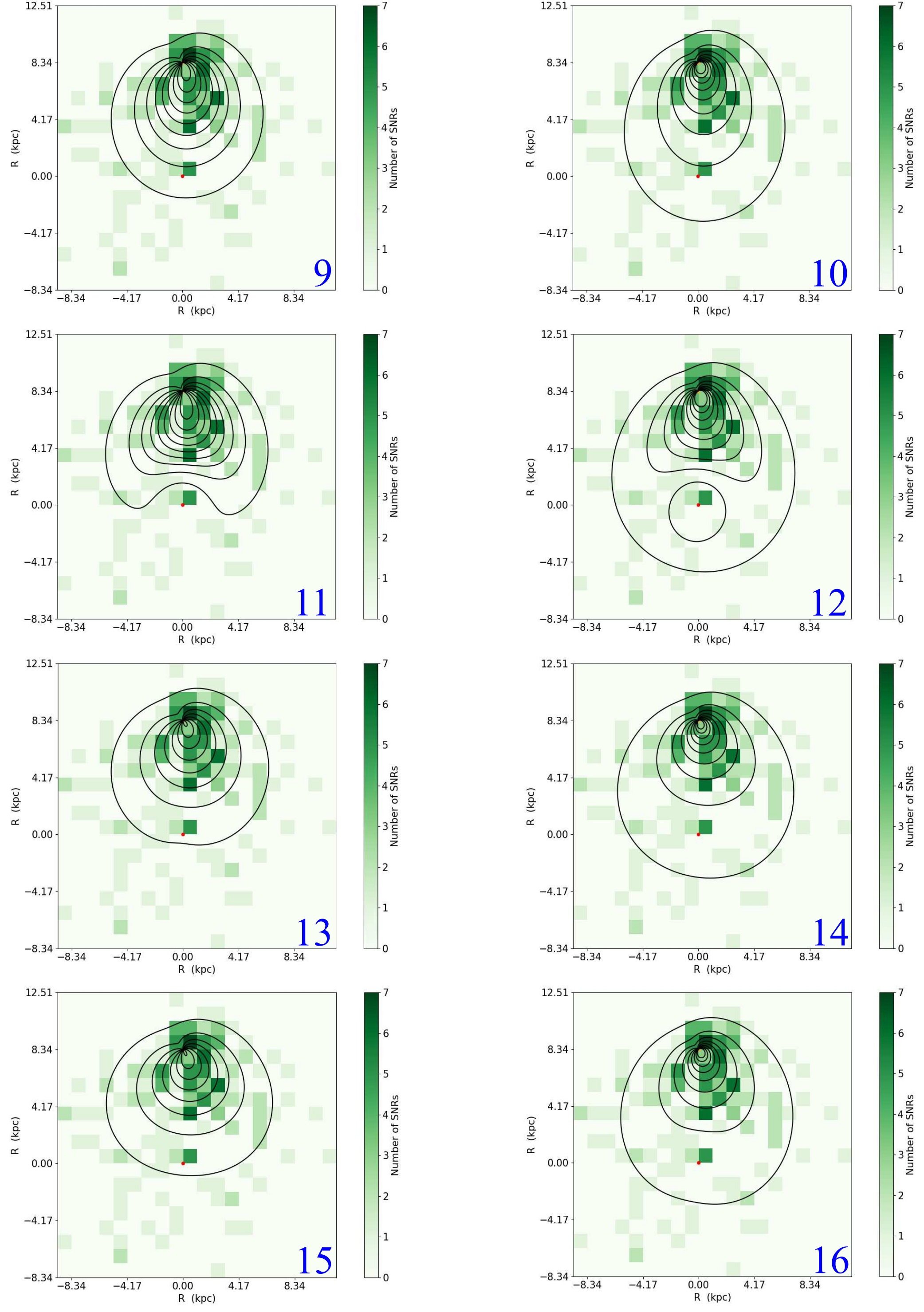}
\caption {The model distributions from Method 2 (product of the SNR surface density functions and selection functions: $A\rho_{\scaleto{SNR}{3pt}}(R)f_{\scaleto{sel}{5pt}}(d,l)$) with black contours at 0.5, 1.5, 2.5,... kpc$^{-2}$.  The numbers in each panel corresponds to the model numbers in Table \ref{tab:3}. The $1.0425 \times 1.0425$ kpc green squares represents the observed number of SNR contained in each area according to the color bars at the left of each panel. The red circle indicates GC and the yellow circle the Sun. The panel numbers correspond to the model numbers given in Table \ref{tab:3}.}
\label{fig:6}
\end{figure*}

\clearpage
\bibliography{RadialDistReferences}{}
\bibliographystyle{aasjournal}



\end{document}